%% file: main_kdd.tex
\documentclass[sigconf,nonacm,screen]{acmart}
\usepackage{xcolor,colortbl,booktabs,multirow,array,pifont,rotating}
\usepackage{booktabs}
\usepackage[table]{xcolor}
\usepackage{tabularx}
\usepackage[most]{tcolorbox}
\tcbuselibrary{skins,breakable}
\usepackage{makecell}
\usepackage{float}
\usepackage[caption=false]{subfig}
\usepackage{listings}
\usepackage{xcolor}
\usepackage{tikz}
\usetikzlibrary{arrows.meta,positioning,decorations.pathreplacing}
\usepackage{amsmath}

\usepackage{amssymb}

\renewcommand\footnotetextcopyrightpermission[1]{}
\makeatletter

\renewcommand{\@fnsymbol}[1]{%
  \ensuremath{%
    \ifcase#1\or \mathsection\or  \dagger\or *\or \ddagger\or \mathsection\or \mathparagraph\or \|\or **\or \dagger\dagger\or \ddagger\ddagger \else\@ctrerr\fi%
  }%
}

\begin{document}

\title{Quantum Circuit Vision: Cost-Aware Evaluation of Visual AI Agents for Quantum Code Generation}

\author{Dongping Liu}
\affiliation{%
  \institution{Amazon Web Services}
  \city{Hong Kong}
  \country{China}}

\authornote{Work done while at Amazon Web Services. Dongping Liu is currently with Tenorshare, Hong Kong, China.}

\authornotemark[2]
\author{Aoyu Zhang}
\affiliation{%
  \institution{Amazon Web Services}
  \city{Beijing}
  \country{China}}

\authornotemark[2]

\author{Luyao Zhang}
\affiliation{%
  \institution{Duke Kunshan University}
  \city{Suzhou}
  \country{China}}
\authornote{The authors are listed in alphabetical order according to last names and, then, first names. \textbf{Acknowledgments}: We thank the anonymous reviewers of the KDD Workshop on Evaluation and Trustworthiness of Agentic AI, KDD-2026, August 9, 2026, Jeju, South Korea for their constructive feedback, which greatly improved the quality and rigor of this work.}

\authornote{The Corresponding author: Email: lz183@duke.edu, Digital Innovation Research Center and Social Science Division, Duke Kunshan University, Address: Duke Avenue No.8, Kunshan, Suzhou, Jiangsu, China, 215316.}

\keywords{Quantum Computing; Multimodal LLMs; Visual Code Generation; Cost-Aware Evaluation; Benchmark}

\begin{abstract}
Can AI agents visually comprehend quantum circuit diagrams and generate verified executable code---and at what cost? We present Quantum Circuit Vision, a cost-aware evaluation framework for multimodal AI agents on quantum circuit visual understanding. We construct a 132-circuit benchmark spanning 13 categories (1--10 qubits) with executable Amazon Braket code and unitary-fidelity verification. Evaluating three frontier Claude-family models at different capability-cost tiers with $n=5$ repeated trials, we find that the mid-tier model (Sonnet~4.6, 1.30$\times$ credits) offers the most favorable balance on the cost--accuracy frontier: 91\% pass rate on the core subset at 18\% of the per-call cost of the strongest model (Opus~4.6), whose accuracy advantage is not statistically significant (paired $t$: $p=0.083$). Logistic regression confirms that circuit depth---not qubit count---is the primary predictor of failure ($p < 0.001$). Chain-of-thought prompting shows no statistically significant effect (all $p > 0.18$, $n=5$), suggesting that visual pattern recognition outweighs explicit reasoning strategy for structurally coupled diagrams. We propose a cascade routing strategy (cheap$\to$expensive models) that achieves 84\% accuracy at 38\% of single-model cost, demonstrating that model routing dominates prompt engineering as a cost lever. We release QCV-Dataset (132 circuits, 5 modalities, 1,931 files) on Hugging Face Hub as an open evaluation infrastructure with structured metadata for discoverability, interoperability, and responsible AI documentation, and all evaluation code, cost logs, and verification scripts on GitHub for full reproducibility.
\end{abstract}

\maketitle

\section{Introduction}

The year 2025 marked a quantum inflection point: the Nobel Prize in Physics recognized experiments with superconducting qubits that enabled macroscopic quantum coherence, while the ACM Turing Award honored the foundations of quantum information theory---including the BB84 protocol that underpins quantum key distribution. The United Nations declared 2025 the International Year of Quantum Science and Technology. Quantum computing has crossed from theoretical promise to engineering reality~\cite{arute2019supremacy,kim2023utility,acharya2023suppressing}. Concurrently, the convergence of artificial intelligence and quantum computing has emerged as a transformative research frontier, with AI techniques advancing challenges across the full quantum stack---from device design and error correction to circuit compilation and verification~\cite{alexeev2025aiquantum}.

Yet a fundamental bottleneck persists: the gap between how quantum algorithms are \emph{communicated} (as circuit diagrams in papers, textbooks, and courses) and how they are \emph{executed} (as code in software development kits (SDKs) such as Amazon Braket~\cite{braket}, Qiskit~\cite{qiskit}, or Cirq). Every quantum circuit published in a research paper must be manually translated into executable code---a tedious, error-prone process requiring both visual interpretation skills and SDK-specific programming expertise. Recent surveys have mapped AI's growing role in quantum preprocessing, including circuit compilation and synthesis~\cite{alexeev2025aiquantum}, yet the specific challenge of \emph{visually interpreting} circuit diagrams and generating verified executable code remains unaddressed. As quantum hardware scales and the volume of published circuits grows exponentially, this manual translation becomes an unsustainable bottleneck for scientific progress.

We argue that \textbf{automating the translation from quantum circuit diagrams to verified executable code is a foundational AI-for-Science challenge}. Inspired by how AlphaFold~\cite{jumper2021alphafold} accelerated structural biology by automating protein structure prediction, we envision that a system reliably reading quantum circuit diagrams and producing verified code could similarly accelerate quantum algorithm development, quantum-safe cryptography research~\cite{fedorov2018quantum,liu2026quantumsafe}, and the broader quantum computing ecosystem.

\textbf{Why quantum circuits, not classical?} Classical integrated circuits have mature electronic design automation (EDA) toolchains developed over 50 years. Quantum circuits have \emph{none of these tools}: every circuit must be manually designed, coded, and verified. Yet the quantum computing market is projected to reach \$450--850B in economic value by 2040~\cite{mckinsey2024quantum}, and the bottleneck is not hardware but the \emph{software toolchain} for designing and verifying quantum algorithms at scale~\cite{preskill2018quantum,cerezo2021variational}.

\textbf{Why autonomous design, not just translation?} Translation (diagram$\to$code) is the first step, but the ultimate value lies in AI systems that can \emph{invent} new quantum circuits---discovering more efficient decompositions, novel error correction codes~\cite{acharya2023suppressing}, or better variational ans\"{a}tze~\cite{cerezo2021variational}.

\textbf{Why cost-aware evaluation?} To our knowledge, no prior work jointly evaluates accuracy, latency, and cost for visual AI agents on scientific diagrams. Deploying AI agents at scale requires understanding not only \emph{what} they can do but \emph{at what cost}. As agentic AI systems are increasingly used for scientific tasks, evaluation frameworks must jointly optimize accuracy, latency, and computational cost---enabling practitioners to select the right model tier for their budget and quality requirements.

In this paper, we take the first step toward this vision by establishing Quantum Circuit Vision (QCV)---the first evaluation framework including multimodal quantum circuit dataset, the cost-aware evaluation methodology, and baseline results that future circuit-discovery systems will build upon.

\section{Related Work}

\subsection{Quantum Circuit Datasets for Machine Learning}

The field of quantum machine learning has historically lacked standardized large-scale datasets, with most experiments relying on custom data processing and evaluation methods rather than shared benchmarks~\cite{placidi2026mnisq}. This gap has motivated a growing body of work on \emph{quantum datasets}---structured data associated with quantum systems that serve as training and evaluation resources for both quantum and classical machine learning models.
\input{tabs/qcv_comparison_table2.tex}
\textbf{NTangled}~\cite{schatzki2021ntangled} represents one of the earliest systematic efforts, introducing entangled quantum states generated by training quantum neural networks to produce states with varying levels of multipartite entanglement. NTangled shares a common motivation with our work in emphasizing quantum-specific datasets for quantum machine learning (QML) advancement. However, it defines the quantum state itself as the dataset, which fundamentally differs from circuit-level datasets: quantum states cannot be directly applied to existing classical ML frameworks without measurement or tomographic reconstruction~\cite{placidi2026mnisq}.

\textbf{QDataSet}~\cite{perrier2022qdataset}, introduced by Perrier, Youssry, and Ferrie in \emph{Scientific Data} (2022), comprises 52 distinct types of data concerning quantum operations for single and two-qubit systems, encompassing both noise-inclusive and noise-free scenarios. Each sub-dataset contains 10,000 data points. However, QDataSet's focus is on the \emph{physical dynamics} of small (1--2 qubit) systems rather than algorithmic circuits.

\textbf{Quantum Federated Data}~\cite{chehimi2022quantum}, proposed by Chehimi and Saad (ICASSP 2022), addresses the distributed learning setting, representing the first quantum federated dataset for distributed quantum computing networks. While the federated setting is important, its focus on distributed learning differs from our emphasis on visual comprehension and code generation.

\textbf{PennyLane Quantum Datasets}~\cite{bergholm2018pennylane} provide curated collections for quantum chemistry and quantum spin systems. However, two limitations are noteworthy: dataset sizes remain relatively small for state-of-the-art ML (typically hundreds to thousands of samples), and the gap between dataset availability and effective quantum/classical ML application remains largely unaddressed.

\textbf{The VQE-Generated Dataset}~\cite{nakayama2025vqe} is the most methodologically similar prior work to MNISQ, comprising quantum circuits labeled by computational task. However, a critical scalability issue limits its applicability: only approximately 300 circuits per label---far below the threshold required for modern deep learning models.

\textbf{MNISQ}~\cite{placidi2026mnisq} (Placidi \textit{et al.}, \emph{Scientific Data} 2026)---a dataset encoding the Modified National Institute of Standards and Technology (MNIST) family of classical image benchmarks into quantum circuits---represents a paradigm shift, containing 4.95 million quantum circuits mapped into 10-qubit systems with up to 100 two-qubit gates. MNISQ baseline experiments show that quantum kernel methods achieve up to 97\% accuracy on multiclass classification, while classical sequence models (S4~\cite{gu2022s4}, Transformers~\cite{vaswani2017attention}, LSTM~\cite{hochreiter1997lstm}) achieve up to 77.78\% on QASM descriptions. While MNISQ studies circuit classification at scale, \emph{QCV-Dataset} addresses a different but complementary question: can AI agents correctly \emph{generate} quantum circuits from visual diagram specifications, and how does structural complexity affect this capability?

\textbf{QCalEval}~\cite{cao2026qcaleval}\footnote{ https://huggingface.co/datasets/nvidia/QCalEval} introduces 243 expert-annotated quantum calibration plots spanning 87 scenario types across 22 experiment families---including Rabi oscillations, DRAG pulse calibration, and single-shot readout GMM clustering---to benchmark vision-language models on quantum experimental data understanding. Each sample is evaluated through a six-part semantic scoring rubric (plot description, experimental conclusion, significance assessment, fit quality, parameter extraction, and actionable recommendation), with ground truth established via human-AI collaborative annotation and expert cross-validation. QCalEval is the first multimodal AI benchmark targeting quantum computing visuals; however, it addresses a fundamentally different task from ours---visual question answering on calibration plots versus circuit diagram-to-code generation---and its dataset is not at the circuit level. Its inclusion in Table~\ref{tab:comparison} reflects its thematic relevance to multimodal AI for quantum error correction experiments, where quality of expert annotation outweighs scale. 

Table~\ref{tab:comparison} situates QCV-Dataset within this landscape along five dimensions spanning dataset governance, ground-truth verification, cost-aware deployment, statistical rigor, and production monitoring. 
\paragraph{Dataset Infrastructure.} Prior quantum datasets, including NTangled, QDataSet, MNISQ, and VQE-generated, provide quantum circuit data at varying scales but lack machine-readable citations (CFF), structured dataset documentation following the Gebru \textit{et al.}~(2021) Datasheets standard~\cite{gebru2021datasheets}, or versioned reproducibility---none combine all three. Visual-to-code benchmarks (VGV, MMCode, Plot2Code, ChartCoder, CODE-VISION) provide no quantum circuit data at all. QCalEval~\cite{cao2026qcaleval} provides multimodal visual input (calibration plot images paired with structured question-answer pairs) but, like the visual-to-code benchmarks, does not contain quantum circuit data. QCV-Dataset is the first to offer all three governance layers together with quantum circuit data. 

\paragraph{Task \& Verification.} Existing benchmarks form two disjoint camps: visual-to-code benchmarks support diagram-to-code generation but rely on functional simulation or surface-level text similarity metrics (BLEU score, pass@k) that accept semantically wrong outputs; quantum verification literature (Burgholzer \& Wille~\cite{burgholzer2022handling}, Viamontes \textit{et al.}~\cite{viamontes2007checking}) provides rigorous unitary equivalence checking but has never been applied to AI agent evaluation. QCalEval~\cite{cao2026qcaleval} bridges neither camp: it evaluates visual comprehension of quantum calibration plots through a semantic scoring rubric (with human-expert-validated ground truth), but does not involve code generation or unitary verification. QCV-Dataset bridges both: it is the \emph{first} benchmark to pair multimodal visual-code generation with unitary-fidelity verification ($F = |\mathrm{Tr}(U_{\text{gt}}^{\dagger} U_{\text{gen}})| / 2^n \geq 0.99$) via a quantum simulator as a domain-expert oracle, establishing an objective, non-LLM-judge ground truth. 

\paragraph{Cost-Aware Deployment.} FrugalGPT~\cite{chen2023frugalgpt} pioneered LLM cascading for cost reduction but lacks per-invocation latency and billing logging, and operates in domains with inherently ambiguous ground truth. QCV-Dataset introduces the first cost-aware evaluation framework for visual AI agents, recording per-call credits, latency, and unitary fidelity, with a cascade routing strategy achieving 84\% accuracy at 38\% of single-model cost---all verified against an objective oracle. 
\paragraph{Statistical Rigor.} No prior work---quantum datasets, visual code benchmarks, cost-aware frameworks, or QCalEval---reports repeated trials with significance testing. QCV-Dataset provides $n=5$ repeated trials per condition, paired $t$-tests across prompting strategies, and logistic regression confirming that circuit depth ($p < 0.001$), not qubit count ($p = 0.20$), predicts comprehension difficulty. 

\paragraph{Trustworthiness \& Monitoring.} QCV-Dataset contributes 268 annotated failure cases across syntax errors, execution errors, and low-fidelity outputs, together with structured CSV logs per invocation---enabling real-time monitoring of deployed visual AI agents. Collectively, these five dimensions demonstrate that QCV-Dataset is the first benchmark to address the full evaluation lifecycle of agentic AI for scientific diagram interpretation.

\subsection{Multimodal AI for Visual Code Generation}
The emergence of multimodal large language models (MLLMs) capable of processing both visual and textual inputs has opened new frontiers in automated code generation from visual specifications.

\textbf{VGV}~\cite{wong2024vgv} (Wong \textit{et al.}, DAC 2024) is the most direct methodological predecessor to our work. VGV demonstrated that MLLMs can generate Verilog hardware description language (HDL) code from classical circuit diagrams, introducing \emph{Thinking Vision} (TV) prompting---a chain-of-thought approach that improved open-source model performance by up to 50 percentage points. Our work extends VGV in several critical directions: (1)~we target quantum rather than classical circuits, finding that standardized gate symbols, strictly left-to-right temporal structure, and finite gate vocabulary yield higher baseline accuracy (78--91\% vs.\ 40--60\% in VGV); (2)~we find that chain-of-thought prompting has \emph{no statistically significant effect} on quantum circuits (all $p > 0.18$, $n=5$), contrasting sharply with VGV's results; (3)~we introduce unitary matrix fidelity verification, which is more rigorous than VGV's functional simulation for hardware circuits.

Several general-domain benchmarks inform our evaluation methodology. \textbf{MMCode}~\cite{li2024mmcode} evaluates MLLMs on code generation from visually rich programming problems. \textbf{Plot2Code}~\cite{wu2025plot2code} provides a benchmark for code generation from scientific plots. \textbf{ChartCoder}~\cite{zhao2025chartcoder} advanced chart-to-code generation through specialized training datasets. \textbf{CODE-VISION}~\cite{wang2025codevision} evaluated MLLMs' logical understanding from visual charts, finding that existing models struggle with information-dense visual inputs---consistent with our observation that circuit depth (structural complexity) predicts failure better than qubit count.

\subsection{Cost-Aware Evaluation and Agentic AI Deployment}

The KDD 2026 Workshop on Evaluation and Trustworthiness of Agentic AI emphasizes evaluating not only what AI agents can accomplish but also the computational cost, latency, and reliability of their deployment. This section connects our cost-aware evaluation framework to broader trends in efficient LLM deployment.

\textbf{FrugalGPT}~\cite{chen2023frugalgpt} (Chen, Zaharia, and Zou, 2023) established the foundational paradigm for cost-aware LLM deployment, introducing LLM cascades that route queries through models of increasing capability. FrugalGPT achieves up to 80\% cost reduction while matching the accuracy of the strongest model alone. Our cascade routing strategy directly builds on this insight: by routing circuits through Haiku (cheapest), then Sonnet, then Opus (most expensive), we achieve 84\% accuracy at 38\% of single-model cost---demonstrating that the FrugalGPT paradigm extends to scientific diagram interpretation tasks.

Subsequent work has refined this approach: cost-aware contrastive routing~\cite{shirkavand2025costaware}, dynamic model cascading~\cite{moslem2026dynamic}, CascadeDebate~\cite{chang2026cascadedebate}, xRouter~\cite{qian2025xrouter}, and CASTER~\cite{liu2026caster} collectively indicate that cost-aware agent orchestration is transitioning from research curiosity to production necessity. \emph{QCV-Dataset} contributes to this ecosystem by providing a scientific-domain benchmark where cost-accuracy tradeoffs can be rigorously quantified through unitary fidelity verification---an objective ground truth absent from many text-generation evaluation scenarios.

\subsection{Quantum Circuit Verification}

Our verification methodology draws on established work in quantum circuit equivalence checking. The gold standard is comparison of unitary matrix representations~\cite{burgholzer2022handling,viamontes2007checking}. Our pipeline computes the full unitary matrix by running all $2^n$ computational basis states through Amazon Braket's LocalSimulator~\cite{braket} and computes fidelity as $F = |\mathrm{Tr}(U_{\text{gt}}^{\dagger} U_{\text{gen}})| / 2^n$. This approach is more rigorous than state-vector comparison on a single input because it verifies correctness across all possible input states simultaneously. The threshold $F \geq 0.99$ is robust: sensitivity analysis shows identical pass rates at $F \geq 0.95$ and $F \geq 0.99$, with a bimodal distribution where circuits are either nearly perfect ($F > 0.999$) or clearly wrong ($F < 0.95$).

\subsection{Positioning and Contributions}

Large language models (LLMs) have shown promise in hardware design automation. Projects such as Chip-Chat~\cite{blocklove2023chipchat}, VeriGen~\cite{thakur2023verigen}, GPT4AIGChip~\cite{fu2023gpt4aigchip}, ChipNemo~\cite{liu2023chipnemo}, and ChatEDA~\cite{he2023chateda} have explored using LLMs to generate HDL code, EDA scripts, and accelerator designs from textual specifications. Multimodal models such as GPT-4~\cite{openai2023gpt4} and LLaVA~\cite{liu2023llava} have further demonstrated strong visual reasoning capabilities across domains including mathematical problem solving~\cite{lu2023mathvista}. Most relevant to our work, VGV~\cite{wong2024vgv} extended this line of research to the visual domain, showing that MLLMs can generate Verilog code from classical circuit diagrams. VGV introduced Thinking Vision (TV) prompting (+50 percentage points (pp) for open-source models).

Quantum circuit diagrams possess properties that make them amenable to visual code generation: quantum gate symbols follow standardized conventions (e.g., $H$ for Hadamard, $\oplus$ for CNOT targets, $\bullet$ for control qubits), a strictly left-to-right temporal structure with no feedback loops, and a finite gate vocabulary. These properties suggest MLLMs may achieve higher accuracy on quantum circuits than on classical ones---a hypothesis we confirm empirically.

\textbf{Our contributions} are:
\begin{enumerate}
    \item We construct \textbf{QCV-Dataset}: 132 quantum circuits across 13 categories (1--10 qubits) with five core modalities per circuit, publicly released with machine-readable citation (CITATION.cff) and dataset documentation following the Gebru \textit{et al.}\ (2021) Datasheets for Datasets standard~\cite{gebru2021datasheets}.
    \item We introduce a \textbf{cost-aware evaluation framework} for visual AI agents, analyzing cost-accuracy tradeoffs across model tiers and proposing a cascade routing strategy that achieves 84\% accuracy at 38\% of single-model cost.
    \item We provide \textbf{statistically rigorous evaluation} with $n=5$ repeated trials, paired $t$-tests, and logistic regression confirming that circuit depth ($p < 0.001$)---not qubit count ($p = 0.20$)---predicts comprehension difficulty.
   \item We release QCV-Dataset (132 circuits, 5+ modalities, 1,931 files) as an \textbf{open} evaluation infrastructure on Hugging Face Hub with structured metadata for discoverability, machine-readable interoperability, and \textbf{responsible AI} documentation covering provenance, limitations, and known biases. All evaluation code, cost logs, and verification scripts are released for full reproducibility on GitHub. \footnote{URLs hidden for anonymous review.}
\end{enumerate}

Our results show that Claude Sonnet 4.6 achieves the highest and most stable accuracy (91.4\%$\pm$5.2\% on the 21-circuit core benchmark under BV), followed by Opus 4.6 (85.7\%$\pm$5.8\%), at only 18\% of Opus's per-call cost. Across the full 132-circuit benchmark, strong models reach 77--78\% and Haiku 4.5 reaches 43--46\%. Chain-of-thought prompting shows no significant effect for any model.

\section{QCV-Dataset}

We construct QCV-Dataset~\ref{fig:workflow}, a multimodal quantum circuit dataset of 132 circuits across 13 categories, with five core data modalities per circuit (images, executable code, simulation results, structured annotations, and failure cases) plus two supplementary modalities for selected circuits (natural language target descriptions and optimization equivalence pairs). We evaluate all 132 circuits across 13 categories. For the 21-circuit core subset (Basic, Intermediate, Advanced), we conduct $n=5$ repeated trials to assess statistical significance and run-to-run variance. Circuit selection follows standard quantum computing curricula~\cite{nielsen2010quantum}. All circuit diagrams are generated programmatically using Qiskit's \texttt{matplotlib} drawer~\cite{qiskit}, ensuring reproducible and unambiguous visual inputs. Ground truth implementations are provided in both Amazon Braket SDK~\cite{braket} and Qiskit.
\input{figures/workflow}

As in Table~\ref{tab:dataset}, the full dataset spans 13 categories: basic gates (5), intermediate algorithms (10), advanced algorithms (6), blockchain protocols (11), gate-type coverage (15), qubit scaling 4--10q (12), classical algorithms (15), variational circuits (10), error correction codes (8), quantum ML (10), blockchain extended (8), visual variants (10), and BTC/blockchain quantum security (12). The 21-circuit core subset used for $n=5$ repeated evaluation is drawn from the first three categories: 1) \textbf{Basic (5 circuits):} Single-gate or single-entangling-gate circuits testing fundamental gate recognition (Hadamard, CNOT, Bell state, GHZ state, Toffoli). 2) \textbf{Intermediate (10 circuits):} Multi-gate circuits requiring understanding of gate composition, control relationships, and parameterized rotations (SWAP decomposition, 2-qubit QFT, teleportation, Deutsch, superdense coding, Grover 2-qubit, parameterized rotation, Fredkin, shift register, phase estimation). 3) \textbf{Advanced (6 circuits):} Complete quantum algorithms involving complex multi-qubit structures (3-qubit QFT, 3-qubit Grover, VQE ansatz, QAOA, quantum walk, Bernstein-Vazirani).

\input{tabs/dataset2}

Each circuit in the full dataset includes: (1)~a PNG diagram image, (2)~executable Braket SDK code, (3)~executable Qiskit code, (4)~simulation results (state vectors from LocalSimulator), (5)~structured bilingual annotations (English/Chinese), and, where applicable, (6)~a natural language target description for circuit synthesis tasks and (7)~optimization equivalence pairs. Additionally, 268 annotated failure cases from our MLLM experiments are included for error analysis research. All structural annotations (qubit count, depth, gate types) are programmatically extracted from executable code; functional descriptions follow standard definitions from original algorithm papers~\cite{nielsen2010quantum}; category assignments follow established benchmark taxonomies~\cite{preskill2018quantum,cerezo2021variational}. Table~\ref{tab:dataset} summarizes the 13 categories.

\subsection{Dataset Publication and Data Governance}

To ensure broad accessibility, reproducibility, and interoperability across the machine learning and quantum computing research communities, we publish QCV-Dataset on \textit{Hugging Face Hub}. The dataset is organized into four structured configurations: \texttt{circuits} (132 records with images, code, annotations, and state vectors), \texttt{experiments} (792 model invocation results with fidelity scores and pass/fail indicators), \texttt{failures} (27 annotated failure cases), and \texttt{equivalences} (12 circuit equivalence pairs). Users can load the dataset programmatically via the \texttt{datasets} library:

\begin{tcolorbox}[
    colback=gray!5, colframe=gray!50!black,
    boxrule=0.4pt, arc=2pt, left=6pt, right=6pt, top=4pt, bottom=4pt
]
\small
\begin{verbatim}
from datasets import load_dataset
circuits = load_dataset("QuantBlockchain/qcv-dataset",
                        "circuits", split="train")
\end{verbatim}
\end{tcolorbox}

\noindent This one-line interface enables immediate access to all 132 circuits with their five core modalities, eliminating the need for manual data parsing or local file management.

\noindent\textbf{Data governance.} Our dataset publication follows a layered metadata governance framework grounded in two complementary sets of principles. The \textbf{FAIR Principles} (Findable, Accessible, Interoperable, Reusable)~\cite{wilkinson2016fair} provide machine-actionable guidelines for scientific data management: data should carry unique persistent identifiers and rich metadata (Findable), be retrievable by both humans and machines (Accessible), use formal, shared knowledge representations (Interoperable), and include clear licenses and provenance documentation (Reusable). The \textbf{CARE Principles} (Collective Benefit, Authority to Control, Responsibility, Ethics)~\cite{carroll2021care} complement FAIR with a people- and purpose-oriented focus, emphasizing that data practices should benefit the communities they affect, that those communities retain authority over their data, that data stewards act transparently and accountably, and that ethical considerations guide data use throughout its lifecycle. We employ three complementary standards, each anchored in specific FAIR and CARE principles:
\textbf{1.Hugging Face Dataset Card (YAML)}: Structured YAML front matter provides task categorization (image-to-text, text-generation, visual-question-answering), language tags (English, Chinese), license information (MIT), and size annotations. This rich metadata enables automatic indexing by the Hugging Face search engine and downstream dataset aggregation tools, directly supporting FAIR's goals of making data \textit{Findable} (through comprehensive, searchable metadata descriptions) and \textit{Reusable} (through explicit licensing for downstream use).
\textbf{2.Croissant JSON-LD~\cite{NEURIPS2024_9547b09b}}: We provide machine-readable Croissant metadata---an emerging industry standard adopted by Google Dataset Search, Kaggle, and OpenML---that represents the dataset as a structured knowledge graph using JSON-LD. This formal, accessible representation ensures data is \textit{Findable} (through globally unique persistent identifiers) and \textit{Interoperable} (through a standardized vocabulary that enables cross-platform discovery and automated integration into ML pipelines without manual schema mapping).
\textbf{3.Croissant-RAI (Responsible AI)~\cite{NEURIPS2024_9547b09b}}: We extend the Croissant schema with the MLCommons RAI vocabulary, documenting data collection methodology (Qiskit programmatic generation + expert bilingual curation), verification protocols (unitary matrix fidelity $\geq 0.99$ on Braket LocalSimulator), known limitations (framework-specific to Braket SDK; simulation-only; EN/CN bilingual coverage), and dataset biases (23.5\% blockchain-relevant circuits that may skew toward cryptographic applications). This addresses CARE's emphasis on \textit{Collective Benefit} (documenting intended use cases for the research community), \textit{Responsibility} (transparent reporting of limitations and biases), and \textit{Ethics} (considering how downstream applications might be affected), while supporting FAIR \textit{Reusability} through detailed provenance documentation that helps future users assess fitness for their purposes.

\textbf{Implementation.} To construct the HF-compatible dataset, we consolidated all five modalities into a unified Apache Parquet format using the \texttt{datasets} library with explicit \texttt{Features} schema declarations. Circuit diagram PNGs were embedded using the \texttt{Image()} feature type, which encodes image bytes directly into the Parquet files---this enables the Hugging Face Dataset Viewer to render circuit previews interactively in the browser. State vectors were stored as variable-length \texttt{Sequence(Value(``float64''))} arrays to handle varying dimensions across qubit counts. We generated Croissant-RAI metadata as a JSON-LD overlay documenting provenance, limitations, and recommended use cases. The dataset was then pushed to the Hub via \texttt{push\_to\_hub()} with named configurations for each split, ensuring clean separation between circuit records and experimental results.

\section{Evaluation Methodology}

Following the approach established in VGV~\cite{wong2024vgv}, we define two visual prompting modes and build an end-to-end verification pipeline tailored to quantum circuits.

\subsection{Visual Modes}

We evaluate two visual prompting strategies adapted from VGV~\cite{wong2024vgv}:

\textbf{Basic Vision (BV)} provides the circuit diagram image with a direct code generation instruction:

\smallskip
\noindent\fbox{\parbox{0.93\columnwidth}{\small\textit{``Please generate Amazon Braket SDK Python code based on this quantum circuit diagram. Requirements: use \texttt{from braket.circuits import Circuit}; output only complete executable Python code; no explanation, no markdown.''}}}
\smallskip

\textbf{Thinking Vision (TV)} adds a chain-of-thought analysis step before code generation:

\smallskip
\noindent\fbox{\parbox{0.93\columnwidth}{\small\textit{``Please first analyze this quantum circuit diagram: 1.~How many qubit lines? What are their labels? 2.~What gates appear from left to right? 3.~Which gates have control qubits? Which line is control, which is target? Then generate Amazon Braket SDK Python code based on your analysis.''}}}
\smallskip

Unlike VGV, which also varied prompt detail levels (Short/Medium/High), we use a single prompt per mode to isolate the effect of chain-of-thought reasoning from prompt informativeness. Formally, given a circuit diagram image $I$ and model $\mathcal{M}$:
\begin{align}
\text{BV}: \quad c &= \mathcal{M}(I,\; p_{\text{bv}}) \\
\text{TV}: \quad (a, c) &= \mathcal{M}(I,\; p_{\text{tv}})
\end{align}
where $c$ is the generated code and $a$ is the intermediate chain-of-thought analysis produced only under TV. Prompts are in Chinese; all evaluated models are multilingual, and preliminary tests showed no meaningful difference between Chinese and English prompts on this task.

\begin{figure*}[t]
\centering
\begin{minipage}[c]{0.68\textwidth}
  \centering
  \subfloat[Basic: Bell State (2q)]{\includegraphics[height=0.9in]{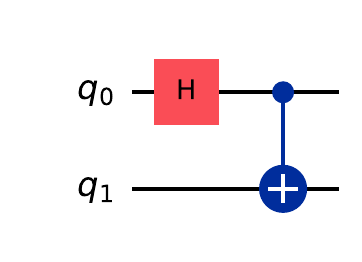}}
  \hfill
  \subfloat[Intermediate: Grover (2q, 9 gates)]{\includegraphics[height=0.9in]{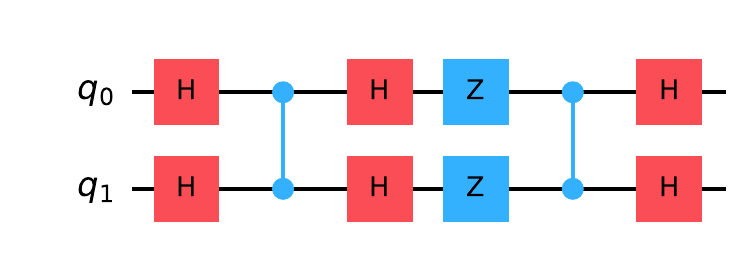}}

  \vspace{0.3em}

  \subfloat[Advanced: QAOA MaxCut (4q, 16 gates) --- all models struggle]{\includegraphics[height=1.3in]{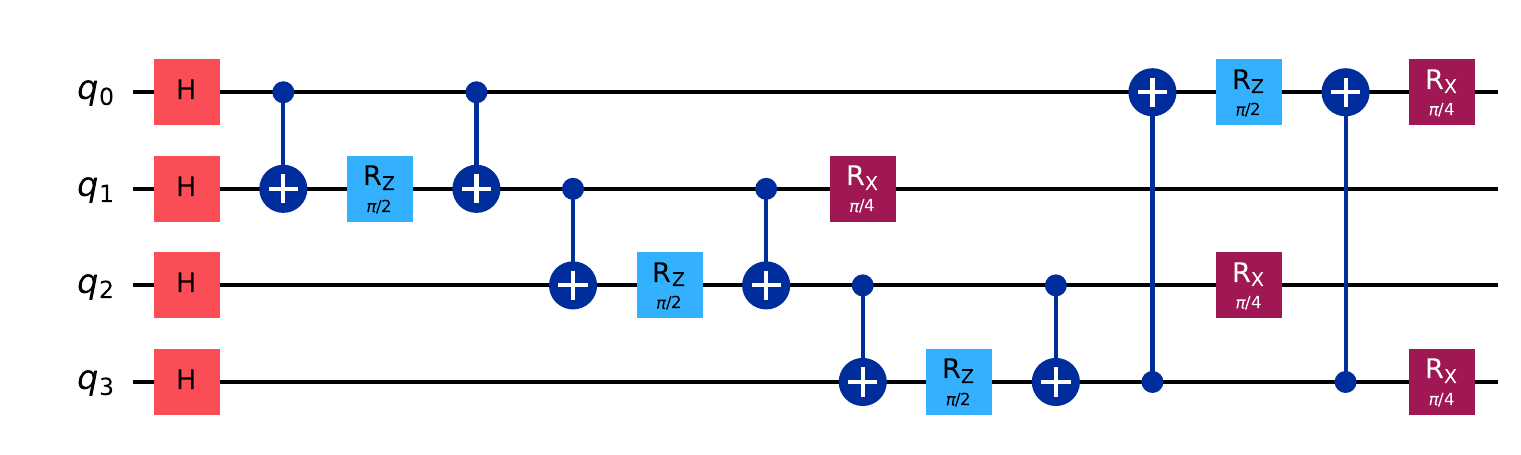}}
\end{minipage}
\hfill
\begin{minipage}[c]{0.30\textwidth}
  \centering
  \subfloat[Blockchain: 8q Consensus --- passes due to regularity]{\includegraphics[height=2.4in]{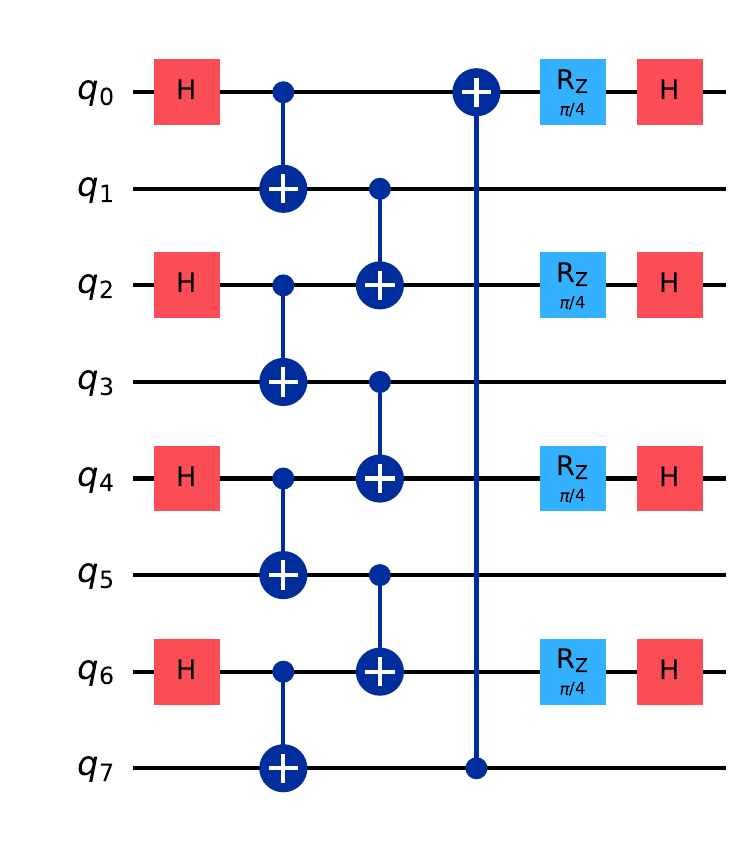}}
\end{minipage}
\caption{Representative circuits. Left column (a--c): increasing complexity from trivial to challenging. Right (d): the largest circuit (8 qubits, 20 gates) passes under BV with perfect fidelity due to its regular repeating structure, while the smaller QAOA (c) fails---demonstrating that structural regularity, not qubit count, determines success.}
\label{fig:circuits}
\end{figure*}

\subsection{Verification Pipeline}

Each model invocation produces a raw text output containing ANSI formatting, tool invocation logs, and the generated code. We first extract the Python code using pattern matching (validated by an independent quality audit; see Appendix~\ref{app:audit}), then verify it through a three-level pipeline:

\begin{enumerate}
    \item \textbf{Syntax check:} The extracted Python code is compiled using \texttt{compile()} to detect syntax errors before execution.
    \item \textbf{Execution check:} The code is executed in a sandboxed namespace with standard imports (\texttt{math}, \texttt{numpy}). We verify that a valid Braket \texttt{Circuit} object is produced and that no runtime exceptions occur (e.g., nonexistent API calls such as \texttt{.crz()} or \texttt{.toffoli()}).
    \item \textbf{Unitary fidelity check:} We compute the full unitary matrix of both the generated circuit and the ground truth by running each of the $2^n$ computational basis states on Amazon Braket's LocalSimulator~\cite{braket} (a free, local state-vector simulator) with \texttt{shots=0}. The fidelity is:
    \begin{equation}
    F = \frac{|\mathrm{Tr}(U_{\text{gt}}^\dagger U_{\text{gen}})|}{d}, \quad d = 2^n
    \end{equation}
    where $U_{\text{gt}}$ and $U_{\text{gen}}$ are the $d \times d$ unitary matrices of the ground truth and generated circuits, respectively. A circuit passes if $F \geq 0.99$.
\end{enumerate}

The unitary-based approach is more rigorous than state-vector comparison on a single input state (e.g., $|0\rangle^{\otimes n}$), as it verifies correctness across \emph{all} possible input states simultaneously. It is also tolerant of global phase differences: two circuits that differ only by a global phase $e^{i\theta}$ will still achieve $F = 1.0$. The threshold $F \geq 0.99$ allows for minor floating-point imprecision while rejecting any structurally incorrect circuit. Sensitivity analysis confirms that conclusions are robust to threshold choice: pass rates are identical at $F \geq 0.95$ and $F \geq 0.99$, and only 1--2 circuits (of 792 invocations) fall between $F = 0.99$ and $F = 0.999$, indicating a bimodal fidelity distribution where circuits are either nearly perfect ($F > 0.999$) or clearly wrong ($F < 0.95$). No cloud quantum hardware is used; all verification runs locally. Because verification operates on the unitary (computed column-wise from state-vector simulation), it is not intrinsically Braket-specific: any SDK exposing a state vector or unitary---Qiskit Aer, Cirq, PennyLane---can be checked by the same oracle, with only a thin execution wrapper differing per SDK. The dataset already ships Qiskit ground truth; empirical cross-SDK verification is future work.

We report the \textbf{pass rate}---the fraction of circuits for which the generated code clears all three levels---per difficulty category and overall, following VGV~\cite{wong2024vgv}:
\begin{equation}
R = \frac{1}{N} \sum_{i=1}^{N} \mathbb{1}\!\left[F\!\left(U_{\text{gt}}^{(i)},\; U_{\text{gen}}^{(i)}\right) \geq 0.99\right]
\end{equation}
where $N$ is the number of circuits in the evaluation set.

\section{Results and Scientific Findings}

\subsection{Experimental Setup}

We evaluate three MLLMs from the Claude family, spanning high-end, mid-tier, and lightweight capability levels. All models are accessed through a unified command-line interface (\texttt{kiro-cli}) in non-interactive mode, ensuring identical execution conditions. Each circuit diagram is provided as a PNG image alongside the BV or TV prompt. All models use default temperature settings with no system prompts beyond the BV/TV templates. The total experiment comprises 132 circuits $\times$ 3 models $\times$ 2 visual modes = 792 invocations. We first report results on the 21-circuit core benchmark (Demo + Intermediate + Advanced) for comparison with prior work, then present the full 132-circuit evaluation across all 13 categories.

All generated code is verified on Amazon Braket's LocalSimulator (a free, local state-vector simulator) using the unitary fidelity pipeline described in Section~3. No cloud quantum hardware is used.

\subsection{Results}
\textbf{Core benchmark (Table~\ref{tab:results}).} With $n=5$ repeated trials on the 21-circuit core benchmark, Sonnet emerges as the most accurate and stable model under BV (91.4\%$\pm$5.2\%), followed closely by Opus (85.7\%$\pm$5.8\%). Chain-of-thought prompting (TV) shows no statistically significant effect for any model: Opus $\Delta=-3.8$pp ($p=0.26$), Sonnet $\Delta=-2.9$pp ($p=0.19$), Haiku $\Delta=+4.8$pp ($p=0.69$). Haiku scores substantially lower (48.6\%$\pm$4.0\% BV), confirming a clear capability gap. Notably, run-to-run variance is substantial ($\pm$4--9pp), demonstrating that single-run evaluations can be misleading.
\input{tabs/result2}
\textbf{Full benchmark (Table~\ref{tab:full-benchmark}).} Scaling to 132 circuits reveals a clear difficulty gradient. Basic circuits (Demo, Gate Coverage) are solved at 80--100\% by all models. Medium circuits (Intermediate, Visual Variants, Qubit Scaling) remain above 90\% for strong models but drop to 30--66\% for Haiku. Hard circuits (Classical Algorithms through BTC/Security) challenge even the strongest models, with pass rates ranging from 41\% to 87\% for Opus. The overall pass rates---Opus 78\%, Sonnet 77\%, Haiku 43\%---establish that these three Claude-family models can reliably interpret standard quantum circuits but face clear limits on complex algorithms and cryptographic protocols. We emphasize that the full 132-circuit results (Table~\ref{tab:full-benchmark}), the cost analysis (Table~\ref{tab:cost}), the failure taxonomy, and the cascade figure are single-run ($n=1$) and should be read as indicative; our significance-tested claims are confined to the $n=5$ core subset.
\input{tabs/full-benchmark2}
Of the 132 circuits, 45 are solved by all six model--mode combinations, while 18 defeat all of them. The 18 ``impossible'' circuits are predominantly complex algorithms (Shor, HHL, QAOA), error correction (surface code), and blockchain security protocols---circuits characterized by irregular multi-qubit entanglement patterns rather than high qubit count alone.

To formally test the ``structural complexity, not qubit count'' hypothesis, we fit logistic regression models predicting majority-pass ($\geq$3/6 configurations pass) from circuit features. Circuit depth alone explains 13.2\% of variance ($p < 0.001$), while qubit count alone explains only 4.6\% ($p = 0.012$). In a joint model, depth remains significant (coef $= -0.25$, $p = 0.001$) while qubit count becomes non-significant (coef $= -0.17$, $p = 0.20$). This confirms that \textbf{circuit depth---a proxy for structural complexity---is the primary predictor of AI comprehension difficulty, not the number of qubits}. Here ``primary predictor'' denotes the strongest among the scalar structural features we tested, not a claim of high absolute explanatory power: depth alone accounts for 13.2\% of deviance, and the remainder reflects category-specific factors (entanglement topology, algorithm family) not captured by scalar counts. We further note that depth and gate count are highly collinear (Pearson $r=0.862$); ``depth'' should thus be read as a proxy for overall structural complexity rather than an isolated cause.

The strong-tier gap widens with difficulty: 27--47pp on Basic, 42pp on BTC/Security. All three models lie on the cost--accuracy Pareto frontier: none is strictly dominated. Haiku minimizes cost, Opus attains the highest full-set accuracy (79.4\% vs.\ Sonnet 77.3\%), and Sonnet occupies a favorable middle, capturing ${\sim}97\%$ of Opus's accuracy at ${\sim}18\%$ of its per-call cost. A per-circuit paired test on the core subset ($n=21$) finds the Sonnet--Opus BV difference (+5.7pp) not statistically significant (paired $t$: $p=0.083$; Wilcoxon: $p=0.057$; 95\% CI $[-0.4,+11.8]$pp), so we do not claim either model dominates the other in accuracy. Which model is ``optimal'' therefore depends on the deployer's stated cost--accuracy preference rather than being fixed by Pareto analysis alone; when both matter, we recommend Sonnet as a sensible default.

\subsection{Finding 1: Chain-of-Thought Has No Significant Effect}

Across the 21-circuit core subset with $n=5$ repeated trials, chain-of-thought prompting (TV) shows no statistically significant effect for any model (Table~\ref{tab:results}). The full 132-circuit benchmark ($n=1$) shows similar trends ($\Delta = -4$, $-3$, $+5$pp; Table~\ref{tab:full-benchmark}) but lacks statistical power for formal testing. For each circuit $i$, we compute the mean pass rate over five runs under each mode, then apply a paired $t$-test:
\begin{equation}
d_i = \bar{r}_i^{\,\text{TV}} - \bar{r}_i^{\,\text{BV}}, \quad
t = \frac{\bar{d}}{s_d / \sqrt{21}}, \quad
p = P(|T_{20}| > |t|)
\end{equation}
where $\bar{r}_i^{\,\text{BV}}$ and $\bar{r}_i^{\,\text{TV}}$ are the mean pass rates for circuit $i$ under BV and TV, respectively. As shown in Table~\ref{tab:results}, no model achieves a statistically significant CoT effect (all $p > 0.18$). The largest effect (+4.8pp for Haiku) has $p = 0.69$, far from significance.

This contrasts with VGV~\cite{wong2024vgv}, where TV improved open-source model performance by up to 50pp on classical circuits. We attribute the difference to quantum circuits' structural properties: the tightly coupled multi-qubit operations (e.g., oracle--diffusion pairs in Grover, cost--mixer layers in QAOA) resist decomposition into sequential reasoning steps. The step-by-step analysis that TV enforces does not help---and may slightly hurt---because the model must ultimately generate code that respects the circuit's holistic structure.

To distinguish absence of evidence from evidence of absence, we add a two-one-sided-tests (TOST) equivalence analysis (equivalence margin $\pm$10pp, per-circuit paired, $n=21$): for Opus ($\Delta=-3.8$pp, TOST $p=0.037$) and Sonnet ($\Delta=-2.9$pp, TOST $p=0.001$), TV and BV are statistically equivalent within $\pm$10pp, whereas for Haiku ($\Delta=+4.8$pp, TOST $p=0.33$) equivalence is not established. We therefore conclude that for the two stronger models chain-of-thought is statistically equivalent to base prompting, rather than universally unhelpful. While individual circuits may benefit from TV (Fig.~\ref{fig:bv_tv_example}), these gains are offset by equal losses on other circuits, resulting in no net effect. The simpler prompting mode achieves equal or better accuracy at lower token cost, and the choice of model (Sonnet vs.\ Haiku: 43pp gap) matters far more than the choice of prompting strategy (BV vs.\ TV: $<$5pp, not significant).

\begin{figure}[!htbp]
\centering
\begin{minipage}[t]{0.48\columnwidth}
\centering\small\textbf{BV Output (Fail)}
\begin{lstlisting}[basicstyle=\ttfamily\scriptsize,frame=single,backgroundcolor=\color{red!8},xleftmargin=0pt,xrightmargin=0pt]
circuit = Circuit()
circuit.toffoli(
  control_qubits=[0, 1],
  target_qubit=2)
\end{lstlisting}
\vspace{-0.5em}
{\scriptsize\color{red} \texttt{.toffoli()} does not exist in Braket SDK}
\end{minipage}
\hfill
\begin{minipage}[t]{0.48\columnwidth}
\centering\small\textbf{TV Output (Pass)}
\begin{lstlisting}[basicstyle=\ttfamily\scriptsize,frame=single,backgroundcolor=\color{green!8},xleftmargin=0pt,xrightmargin=0pt]
circuit = Circuit()
circuit.ccnot(0, 1, 2)

\end{lstlisting}
\vspace{-0.5em}
{\scriptsize\color{green!50!black} Correct API: \texttt{.ccnot()}, fidelity = 1.0}
\end{minipage}
\caption{Haiku on the Toffoli circuit: BV uses a nonexistent API, while TV's analysis leads to the correct call. Although CoT shows no significant aggregate effect ($p=0.69$), individual cases like this illustrate how structured analysis can resolve specific API selection errors.}
\label{fig:bv_tv_example}
\end{figure}

\subsection{Finding 2: Failure Mode Taxonomy}

Across 792 invocations, we observe 524 passes and 268 failures. Failure modes stratify sharply by model capability:

\textbf{Opus} (28 BV failures): dominated by \emph{low fidelity} (22/28)---the model generates syntactically valid, executable code that produces the wrong unitary. This indicates correct visual interpretation but incorrect gate parameters or ordering. Only 3 failures are syntax errors and 3 are execution errors.

\textbf{Sonnet} (30 BV failures): a mix of \emph{low fidelity} (13/30) and \emph{execution errors} (17/30). The execution errors typically involve nonexistent Braket API calls (e.g., \texttt{.cp()}, \texttt{.crz()})---correct quantum operations expressed in the wrong SDK dialect.

\textbf{Haiku} (75 BV failures): dominated by \emph{low fidelity} (54/75), with substantial \emph{execution errors} (21/75). Unlike Opus, Haiku's fidelity failures often reflect fundamental visual misinterpretation---reversed CNOT directions, omitted gates, or incorrect qubit assignments---rather than subtle parameter errors.

\subsection{Finding 3: Cost-Accuracy Tradeoff}

Deploying AI agents for scientific tasks requires evaluating not only accuracy but also computational cost. We analyze the cost-accuracy relationship across five dimensions.

\textbf{Per-model cost.} Table~\ref{tab:cost} reports the per-invocation cost (in platform credits), latency, and cost-efficiency for each model under BV mode. We define cost-per-correct-circuit as:
\begin{equation}
C_{\text{eff}} = \frac{\sum_{i=1}^{N} c_i}{\sum_{i=1}^{N} \mathbb{1}[F_i \geq 0.99]}
\end{equation}
where $c_i$ is the credit cost of invocation $i$.

\input{tabs/cost2}

\textbf{Diminishing returns.} The marginal cost (MC) of accuracy improvement exhibits extreme diminishing returns. Upgrading from Haiku to Sonnet costs 0.079 additional credits per call for +34pp accuracy (MC = 0.002 credits/pp)---an excellent investment. Upgrading from Sonnet to Opus costs 0.508 additional credits for only +1pp (MC = 0.508 credits/pp)---a 254$\times$ higher marginal rate. This follows a classic diminishing-returns curve where the last percentage points of accuracy are disproportionately expensive.

\textbf{Cost scales with difficulty.} Opus's per-call cost increases 78\% from Basic circuits (0.41 credits) to Hard circuits (0.73 credits), reflecting longer outputs and more tool invocations for complex circuits. Sonnet shows a smaller increase (39\%), while Haiku remains nearly flat. Complex circuits thus impose a \emph{double penalty}: lower accuracy \emph{and} higher cost.

\textbf{Cascade routing strategy.} An agentic deployment can reduce cost by routing circuits through models of increasing capability. Using our actual experimental data:
\begin{enumerate}
\item Run all 132 circuits with Haiku: 57 pass (43\%), cost = 4.1 credits.
\item Re-run 75 failures with Sonnet: 45 pass (60\% rescue rate), cost = 8.3 credits.
\item Re-run 30 remaining with Opus: 10 pass (33\% rescue rate), cost = 18.5 credits.
\end{enumerate}
This cascade achieves \textbf{112/132 = 84\% pass rate at 30.9 credits} (38\% of Opus-only cost of 81.6 credits), demonstrating that \textbf{intelligent routing is a key design dimension for cost-effective agentic AI systems}.

\textbf{Return on investment (ROI) vs.\ human translation.} Manual quantum circuit translation requires an estimated 1--4 hours per circuit by a trained researcher (based on author experience and informal surveys of quantum computing labs). At an estimated \$50--200/circuit for human labor (assuming \$50--100/hour researcher time), even the most expensive model (Opus: approximately \$0.78/circuit at 78\% accuracy) represents a $>$100$\times$ cost reduction. A hybrid workflow---AI generates, human verifies and fixes the 22\% failures---reduces per-circuit cost from \$100 to approximately \$23 (77\% savings), making large-scale circuit translation economically viable for the first time.


\paragraph{Comparison with VGV.}
Our methodology extends VGV~\cite{wong2024vgv} from classical (Verilog) circuit diagram-to-code generation. Three key differences emerge: (1)~quantum circuits yield higher baseline accuracy (78--91\% vs.\ 40--60\%) due to standardized gate symbols, strict left-to-right temporal structure, and finite gate vocabulary---properties absent in classical microarchitecture diagrams; (2)~chain-of-thought effects diverge: TV improved VGV's open-source models by up to +50\,pp, whereas we find no statistically significant effect on quantum circuits (all $p > 0.18$, $n = 5$), suggesting tightly-coupled multi-qubit structures resist step-by-step decomposition; (3)~failure modes differ: VGV's failures are predominantly syntax errors in hardware description language (HDL) generation, while ours are dominated by low-fidelity outputs---syntactically valid code implementing the wrong unitary---indicating that semantic correctness, not syntactic fluency, is the core challenge.
\subsection{Application: Quantum-Safe Blockchain}

As a domain-specific case study, we evaluate 11 blockchain-relevant circuits (2--8 qubits) spanning quantum threats, post-quantum defenses, and quantum infrastructure. Detailed results are in Appendix~\ref{app:blockchain}. Key finding: Sonnet + BV (9/11) outperforms Opus BV (7/11) on these circuits, reinforcing that the mid-tier model avoids over-analysis errors on circuits with parallel gate layers.

\section{Discussion and Future Work}

Our results establish that multimodal AI can reliably interpret quantum circuit diagrams today (77--91\% accuracy depending on complexity). This positions QCV-Dataset as the foundation for a three-level research trajectory toward autonomous quantum circuit design: \textbf{Level 1 --- Understanding (this paper):} AI reads circuit diagrams and generates verified executable code. We have demonstrated feasibility on 132 circuits across 13 categories with $n=5$ repeated trials on a core subset, finding that structural complexity determines success and that chain-of-thought prompting does not significantly improve performance. The immediate next step is expanding evaluation to non-Claude models (GPT-4o, Gemini); \textbf{Level 2 --- Optimization (near-term):} AI improves existing circuits. Using the equivalence pairs and target descriptions in QCV-Dataset, future systems can learn to suggest more efficient gate decompositions, reduce circuit depth, or adapt circuits to hardware constraints---functioning as a quantum protocol design co-pilot; \textbf{Level 3 --- Invention (long-term):} AI designs novel quantum circuits from specifications, generating, verifying, and iterating autonomously. As a proposed design (not evaluated in this paper), we outline a Hybrid Agent architecture (Appendix~\ref{app:agent}) that would combine QCV-Dataset as a retrieval-augmented generation (RAG) knowledge base, the unitary fidelity pipeline as a verification oracle, and the 268 annotated failure cases as negative examples. This closed-loop design-verify-refine cycle mirrors how human quantum engineers work, but operates at machine speed.

These results also clarify how task structure mediates prompting effectiveness. The finding that CoT has no significant effect on quantum circuit understanding---in contrast to its strong benefits for classical circuits~\cite{wong2024vgv}---suggests that structured reasoning strategies must be matched to task structure. For tasks requiring holistic visual pattern matching over tightly coupled components, step-by-step decomposition may be neither helpful nor harmful.

\textbf{Implications for agentic AI evaluation}: The cost-aware framework generalizes to any scientific diagram domain with verifiable ground truth. Any domain where AI agents perform scientific diagram interpretation (e.g., chemical structures, biological pathways, engineering schematics) faces the same cost-accuracy tradeoff. The cascade routing strategy and diminishing-returns analysis provide a template for evaluating agentic systems in cost-constrained deployments.

\textbf{Limitations and Future Directions}:
\textbf{1.~Model coverage.} We deliberately hold the vendor fixed to isolate the capability--cost tier effect within one model family, evaluating three Claude-family tiers via \texttt{kiro-cli}; other models on the same gateway (DeepSeek, MiniMax, GLM) lack multimodal image input. The pipeline is model-agnostic---any model with image input and code output can be plugged in---so cross-vendor evaluation (GPT-4o, Gemini, Qwen-VL) is a natural direction for future work. \textbf{2. Visual diversity and verification portability.} All circuit diagrams are rendered by Qiskit's matplotlib drawer; real-world circuits vary widely (TikZ, Quirk, hand-drawn). Our Visual Variants category (H) tests layout robustness with 90--100\% pass rates for strong models, but full heterogeneity testing across renderers remains future work. While our unitary-fidelity oracle is not intrinsically Braket-specific (it operates on the circuit unitary computed via simulation), in this work we only ran the end-to-end verification pipeline through Braket's execution wrapper; empirical cross-SDK verification (e.g., Qiskit Aer, Cirq, PennyLane) is future work. \textbf{3. Beyond translation.} This paper addresses Level~1---visual comprehension and verified code generation. The QCV-Dataset already contains equivalence pairs and natural language targets to support Level~2 (optimization) and Level~3 (invention). Each repeated trial is a full paid LLM invocation ($n=5$ over all 132 circuits $\times$3 models$\times$2 modes is ${\sim}3{,}960$ calls), so we prioritized statistical depth on a representative 21-circuit core; broader repeated-trial coverage is a natural direction for future work. More broadly, the cost-aware methodology generalizes to any scientific diagram domain with verifiable ground truth---chemical structures, biological pathways, engineering schematics---positioning this work as foundational infrastructure for multimodal AI in science.


\bibliographystyle{ACM-Reference-Format}
\bibliography{references}

\appendix

\section{Blockchain Circuit Diagrams}
\label{app:blockchain}

\begin{figure*}[h]
\centering
\subfloat[QRNG (4q)]{\includegraphics[height=1.1in]{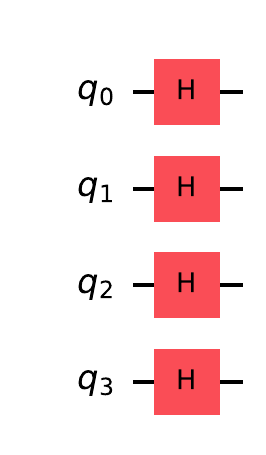}}
\hfill
\subfloat[BV Oracle (3q)]{\includegraphics[height=1.1in]{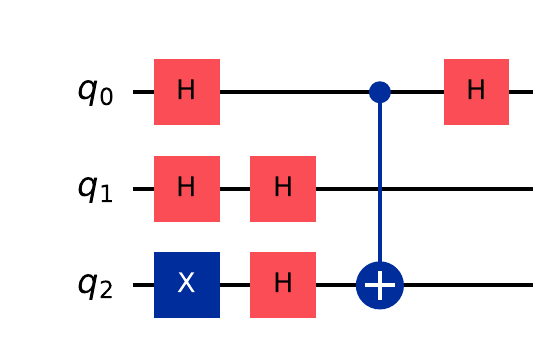}}
\hfill
\subfloat[Grover Search (2q)]{\includegraphics[height=1.1in]{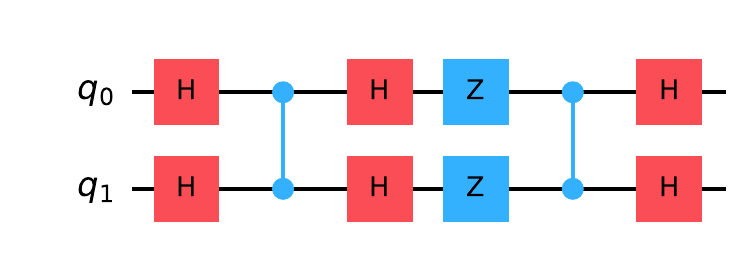}}
\hfill
\subfloat[BB84 QKD (4q)]{\includegraphics[height=1.1in]{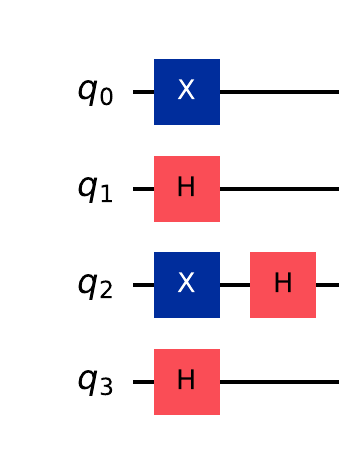}}
\hfill
\subfloat[Lattice PQC (5q)]{\includegraphics[height=1.1in]{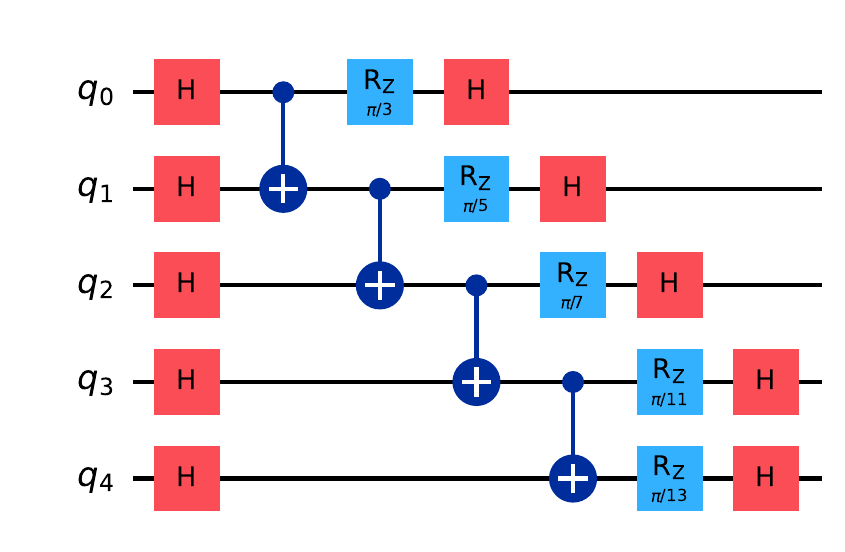}}
\hfill
\subfloat[Quantum Secret Sharing (4q)]{\includegraphics[height=1.1in]{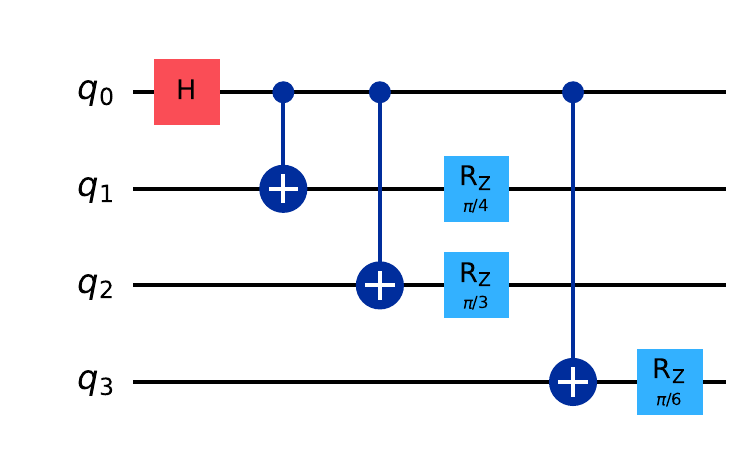}}
\subfloat[Coin Flip (3q)]{\includegraphics[height=1.1in]{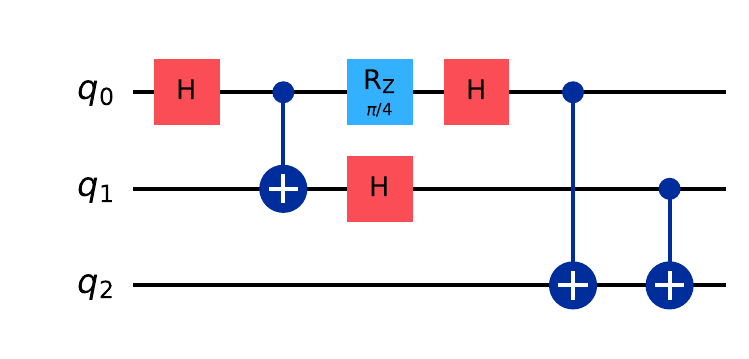}}
\hfill
\subfloat[Oblivious Transfer (5q)]{\includegraphics[height=1.1in]{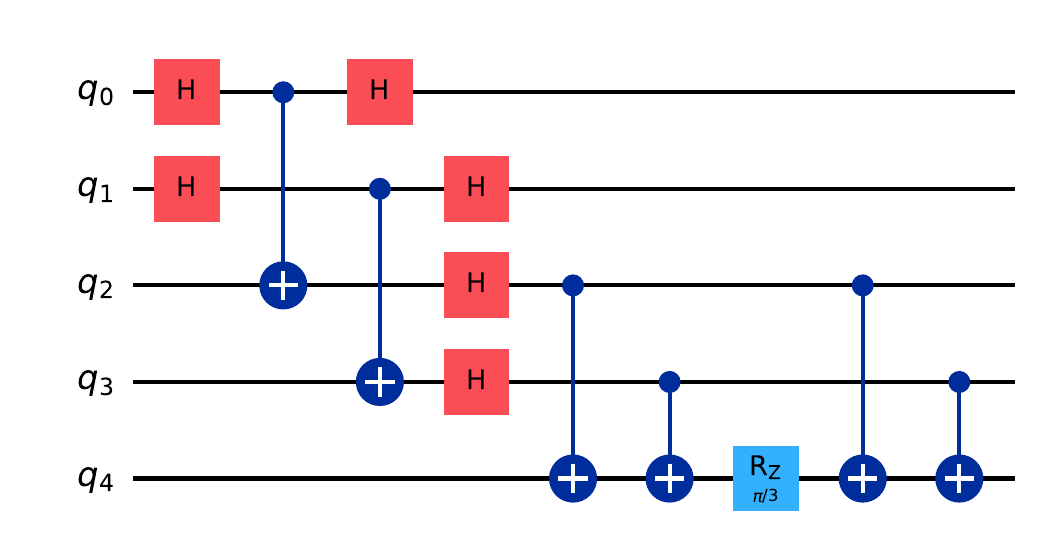}}
\hfill
\subfloat[Quantum VRF (6q)]{\includegraphics[height=1.1in]{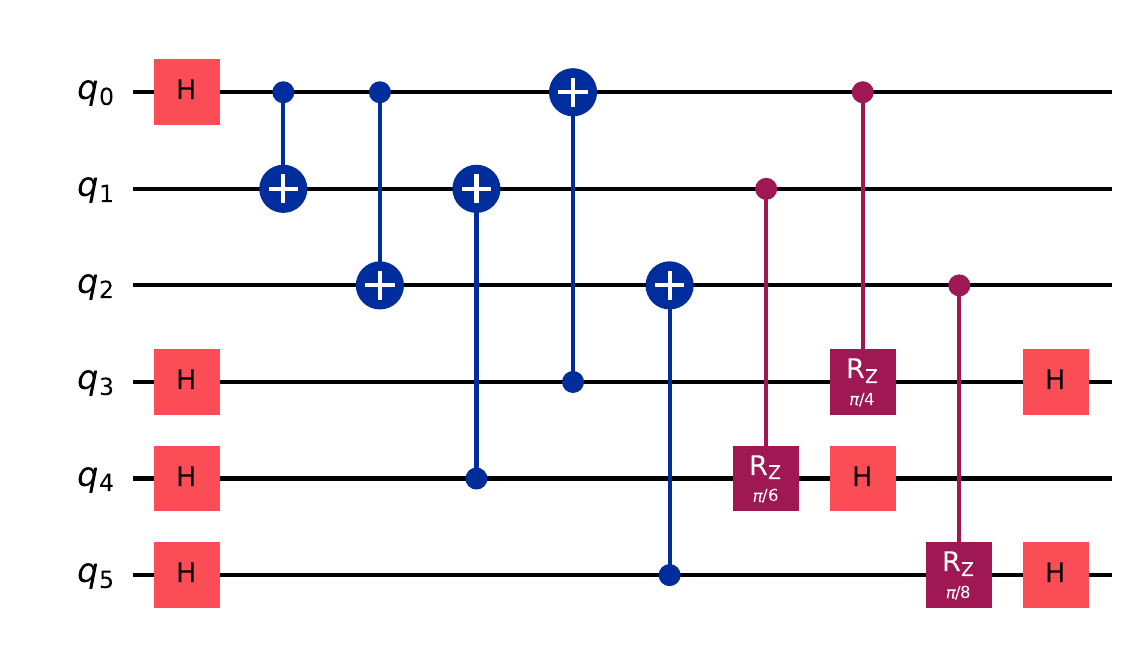}}
\hfill
\subfloat[Digital Signature (7q)]{\includegraphics[height=1.1in]{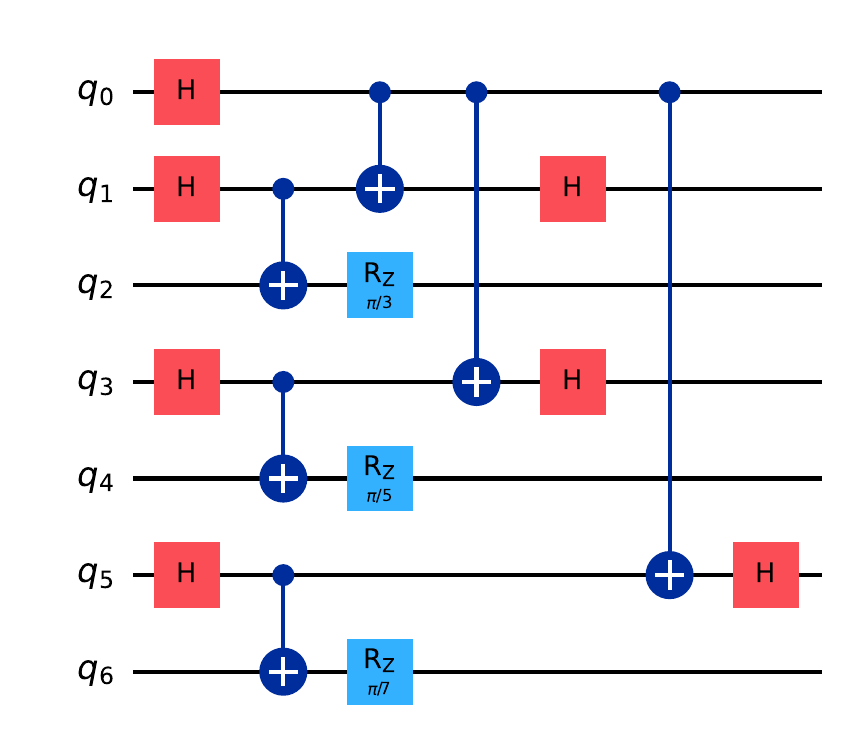}}
\hfill
\subfloat[Consensus Protocol (8q)]{\includegraphics[height=1.1in]{figures/blockchain_consensus.pdf}}
\caption{Eleven blockchain-relevant quantum circuits (2--8 qubits) spanning consensus, cryptography, threat modeling, key distribution, secret sharing, fair protocols, private exchange, verifiable randomness, digital signatures, and validator agreement.}
\label{fig:blockchain}
\end{figure*}

Figure~\ref{fig:blockchain} shows the eleven blockchain-relevant quantum circuits evaluated in Section~4. Circuits range from 2 qubits (Grover Search) to 8 qubits (Consensus Protocol), spanning six blockchain security functions: consensus randomness, cryptographic key analysis, quantum key distribution, secret sharing, fair protocols, and post-quantum defense.

\begin{table}[h]
\centering
\caption{Blockchain circuit results. R=Regular, I=Irregular, P=Parallel. \checkmark=pass ($F\geq0.99$).}
\label{tab:blockchain}
\scriptsize
\begin{tabular}{@{}llcclccc@{}}
\toprule
Circuit & Qubits & Gates & Struct. & Opus BV & Opus TV & Son.\ BV \\
\midrule
QRNG & 4 & 4 & R & \checkmark & \checkmark & \checkmark \\
BV Oracle & 3 & 5 & R & \checkmark & \checkmark & \checkmark \\
Grover & 2 & 9 & I & 0.50 & 0.25 & \checkmark \\
BB84 & 4 & 4 & R & \checkmark & \checkmark & \checkmark \\
Lattice PQC & 5 & 15 & P & 0.00 & 0.50 & \checkmark \\
QSS & 4 & 7 & R & \checkmark & \checkmark & \checkmark \\
Coin Flip & 3 & 8 & R & \checkmark & \checkmark & \checkmark \\
Oblivious Xfer & 5 & 12 & I & \checkmark & \checkmark & \checkmark \\
VRF & 6 & 15 & I & 0.24 & 0.00 & 0.00 \\
QDS & 7 & 16 & I & 0.25 & 0.00 & 0.50 \\
Consensus & 8 & 20 & R & \checkmark & 0.00 & \checkmark \\
\midrule
\textbf{Pass} & & & & \textbf{7/11} & \textbf{6/11} & \textbf{9/11} \\
\bottomrule
\end{tabular}
\end{table}

\section{Circuits That Defeat All Models}
\label{app:failures}

Table~\ref{tab:impossible} lists the 18 circuits that failed across all six model--mode combinations. The dominant failure mode is low fidelity (syntactically valid code that produces the wrong unitary), indicating that models can parse these circuits visually but cannot correctly reconstruct their complex gate sequences. The circuits share two structural properties: irregular multi-qubit entanglement patterns and algorithm-specific gate orderings that do not follow repeating templates.

\begin{table}[h]
\centering
\caption{The 18 circuits that defeat all models ($0/6$ pass). Dominant error: LF = low fidelity, EX = execution error, SY = syntax error.}
\label{tab:impossible}
\small
\begin{tabular}{@{}llc@{}}
\toprule
Circuit & Category & Dominant Error \\
\midrule
A10\_controlled\_rx & Gate Coverage & LF (0.50--0.93) \\
A11\_ccz & Gate Coverage & LF / EX \\
inter\_10\_phase\_est & Intermediate & EX (.crz API) \\
adv\_04\_qaoa & Advanced & LF (0.06--0.50) \\
C05\_shor\_period\_4 & Classical Algo. & LF (0.25--0.63) \\
C08\_hhl\_3 & Classical Algo. & EX (.cp API) \\
C10\_w\_state\_3 & Classical Algo. & SY / EX \\
C11\_w\_state\_4 & Classical Algo. & SY / EX \\
D04\_qaoa\_2layer & Variational & LF (0.02--0.36) \\
E05\_surface\_code & Error Correction & LF (0.003--0.13) \\
G05\_quantum\_auction & Blockchain Ext. & LF (0.13--0.75) \\
I01\_shor\_ecc\_6 & BTC/Security & LF / EX \\
I03\_grover\_aes\_5 & BTC/Security & LF (0.00--0.25) \\
I09\_dilithium\_sign & BTC/Security & LF (0.25--0.50) \\
I10\_pow\_speedup\_4 & BTC/Security & LF (0.00--0.50) \\
I12\_sphincs\_hash\_7 & BTC/Security & LF (0.06--0.38) \\
blockchain\_qds & Blockchain & LF / SY \\
blockchain\_vrf & Blockchain & LF / EX \\
\bottomrule
\end{tabular}
\end{table}

\section{Code Extraction Quality Audit}
\label{app:audit}

To ensure the reliability of our automated code extraction pipeline (which parses model outputs containing ANSI escape codes, diff-style formatting, and tool invocation logs), we performed an independent quality audit. Each extracted code snippet was checked for: (1)~presence of the \texttt{from braket} import, (2)~presence of a \texttt{Circuit()} constructor call, (3)~balanced parentheses, (4)~absence of diff-line residue or tool output contamination. On the full 132-circuit benchmark ($n=1$, 792 files), 792/792 (100\%) passed all checks. On the $n=5$ repeated trials (630 files), 628/630 (99.7\%) passed; the two flagged files (Opus BV, inter\_02\_qft2, runs 3--4) had missing import lines and were counted as failures in our analysis.

\section{Hybrid Agent Architecture}
\label{app:agent}

Figure~\ref{fig:agent} shows the proposed Hybrid Agent for autonomous quantum circuit design. The agent uses QCV-Dataset as a RAG knowledge base, the unitary fidelity pipeline as a verification oracle, and the 268 annotated failure cases as negative examples. This architecture is proposed but not yet implemented.

\begin{figure}[H]
\centering
\begin{tikzpicture}[node distance=0.4cm, every node/.style={font=\small},
  box/.style={draw, rounded corners, minimum height=0.7cm, minimum width=2.4cm, align=center, fill=blue!8},
  dbox/.style={draw, rounded corners, minimum height=0.7cm, minimum width=2.4cm, align=center, fill=orange!10},
  vbox/.style={draw, rounded corners, minimum height=0.7cm, minimum width=2.4cm, align=center, fill=green!8},
  arr/.style={-{Stealth[length=2mm]}, thick},
  darr/.style={-{Stealth[length=2mm]}, thick, dashed}]
\node[box] (user) at (0,5.5) {User Target\\[-2pt]\scriptsize``Design 5q QKD''};
\node[dbox] (rag) at (0,4.0) {RAG Retrieval\\[-2pt]\scriptsize QCV-Dataset};
\node[box] (llm) at (0,2.5) {MLLM Agent\\[-2pt]\scriptsize Generate Circuit};
\node[vbox] (verify) at (0,1.0) {Verification\\[-2pt]\scriptsize Syntax$\to$Exec$\to$Fidelity};
\node[draw, rounded corners, fill=red!10, minimum height=0.6cm, minimum width=1.5cm] (fail) at (2.8,1.8) {\scriptsize Fail + Error};
\node[draw, rounded corners, fill=green!20, minimum height=0.6cm, minimum width=1.5cm] (pass) at (0,-0.3) {\scriptsize Pass $\to$ Output};
\draw[arr] (user) -- (rag);
\draw[arr] (rag) -- node[right, font=\scriptsize] {similar circuits} (llm);
\draw[arr] (llm) -- node[right, font=\scriptsize] {candidate code} (verify);
\draw[arr] (verify) -- (pass);
\draw[arr] (verify.east) -- (fail.south west);
\draw[darr] (fail.north) |- node[above, font=\scriptsize] {feedback loop} (llm.east);
\end{tikzpicture}
\caption{Proposed Hybrid Agent architecture (not yet implemented).}
\label{fig:agent}
\end{figure}

\end{document}

%% file: tabs/qcv_comparison_table2.tex

\definecolor{qcv-infra}{RGB}{214, 232, 255}
\definecolor{qcv-task}{RGB}{204, 242, 220}
\definecolor{qcv-cost}{RGB}{255, 236, 204}
\definecolor{qcv-stat}{RGB}{232, 218, 255}
\definecolor{qcv-trust}{RGB}{255, 214, 224}
\definecolor{qcv-ok}{RGB}{21, 128, 61}
\definecolor{qcv-no}{RGB}{170, 170, 180}
\definecolor{qcv-part}{RGB}{190, 120, 0}
\definecolor{qcv-qcvbg}{RGB}{235, 255, 240}

\providecommand{\cmark}{}
\providecommand{\xmark}{}
\providecommand{\pmark}{}
\renewcommand{\cmark}{\textcolor{qcv-ok}{\ding{51}}}
\renewcommand{\xmark}{\textcolor{qcv-no}{\ding{55}}}
\renewcommand{\pmark}{\textcolor{qcv-part}{\ding{108}}}

\newcolumntype{C}{>{\centering\arraybackslash}p{0.65cm}}
\newcolumntype{L}{>{\raggedright\arraybackslash}p{3.4cm}}
\newcolumntype{T}{>{\raggedright\arraybackslash}p{2.0cm}}

\begin{table*}[t]
\caption{Comparative analysis of QCV-Dataset against existing quantum datasets,
visual code-generation benchmarks, and cost-aware AI evaluation frameworks. Dimensions organized by evaluation taxonomy.}
\label{tab:comparison}
\footnotesize
\setlength{\tabcolsep}{3.5pt}
\renewcommand{\arraystretch}{1.12}
\centering
\begin{tabular}{@{}T L CCCCCCCCCC >{\columncolor{qcv-qcvbg}}c@{}}
\toprule[1.2pt]
\textbf{Taxonomy} & \textbf{Dimension}
& \rotatebox{90}{\textbf{NTangled}}
& \rotatebox{90}{\textbf{QDataSet}}
& \rotatebox{90}{\textbf{Q.\ Fed.}}
& \rotatebox{90}{\textbf{PennyLane}}
& \rotatebox{90}{\textbf{VQE-gen.}}
& \rotatebox{90}{\textbf{MNISQ}}
& \rotatebox{90}{\textbf{QCalEval}}
& \rotatebox{90}{\textbf{VGV}}
& \rotatebox{90}{\textbf{Vis.$\to$Code}}
& \rotatebox{90}{\textbf{FrugalGPT}}
& \cellcolor{white}\textbf{QCV (Ours)}
\\ \midrule[0.8pt]

\cellcolor{qcv-infra}\textbf{Dataset Infra.}
& Scale (samples)
& $\sim$10K & 52$\times$10K & 1,500 & $\sim$100s & $\sim$300 & \textbf{4.95M} & 243 & 59 & 1K--10K & N/A & 132 \\
\cellcolor{qcv-infra}
& Quantum circuit data
& \cmark & \cmark & \cmark & \cmark & \cmark & \cmark & \xmark & \xmark & \xmark & \xmark & \cmark \\
\cellcolor{qcv-infra}
& Multimodal visual input
& \xmark & \xmark & \xmark & \xmark & \xmark & \xmark & \cmark & \cmark & \cmark & \xmark & \cmark \\
\cellcolor{qcv-infra}
& Datasheets for Datasets
& \xmark & \xmark & \xmark & \xmark & \xmark & \xmark & \xmark & \xmark & \xmark & \xmark & \cmark \\
\cellcolor{qcv-infra}
& Machine-readable citation
& \xmark & \xmark & \xmark & \xmark & \xmark & \xmark & \xmark & \xmark & \xmark & \xmark & \cmark \\
\cellcolor{qcv-infra}
& Versioned reproducibility
& \xmark & \xmark & \xmark & \xmark & \xmark & \cmark & \xmark & \xmark & \pmark & \xmark & \cmark \\

\midrule[0.3pt]

\cellcolor{qcv-task}\textbf{Task \& Verify}
& Visual $\to$ executable code
& \xmark & \xmark & \xmark & \xmark & \xmark & \xmark & \xmark & \cmark$^\dagger$ & \cmark$^\ddagger$ & \xmark & \cmark \\
\cellcolor{qcv-task}
& Objective GT verification
& \xmark & \xmark & \xmark & \xmark & \xmark & \xmark & \cmark & \xmark & \xmark & \xmark & \cmark$^\star$ \\
\cellcolor{qcv-task}
& Domain-expert oracle
& \xmark & \xmark & \xmark & \xmark & \xmark & \xmark & \cmark & \xmark & \xmark & \xmark & \cmark$^\diamond$ \\

\midrule[0.3pt]

\cellcolor{qcv-cost}\textbf{Cost Deploy.}
& Cost-aware per-invocation
& \xmark & \xmark & \xmark & \xmark & \xmark & \xmark & \xmark & \xmark & \xmark & \cmark$^\S$ & \cmark \\
\cellcolor{qcv-cost}
& Latency + billing logging
& \xmark & \xmark & \xmark & \xmark & \xmark & \xmark & \xmark & \xmark & \xmark & \xmark & \cmark \\
\cellcolor{qcv-cost}
& Cascade / compound AI
& \xmark & \xmark & \xmark & \xmark & \xmark & \xmark & \xmark & \xmark & \xmark & \cmark & \cmark \\
\cellcolor{qcv-cost}
& Diminishing-returns analysis
& \xmark & \xmark & \xmark & \xmark & \xmark & \xmark & \xmark & \xmark & \xmark & \cmark & \cmark \\

\midrule[0.3pt]

\cellcolor{qcv-stat}\textbf{Stat. Rigor}
& Repeated trials ($n{\geq}5$)
& \xmark & \xmark & \xmark & \xmark & \xmark & \xmark & \xmark & \xmark & \xmark & \xmark & \cmark \\
\cellcolor{qcv-stat}
& Paired significance testing
& \xmark & \xmark & \xmark & \xmark & \xmark & \xmark & \xmark & \xmark & \xmark & \xmark & \cmark \\

\midrule[0.3pt]

\cellcolor{qcv-trust}\textbf{Trust \& Monit.}
& Failure taxonomy (268 cases)
& \xmark & \xmark & \xmark & \xmark & \xmark & \xmark & \xmark & \xmark & \xmark & \xmark & \cmark \\
\cellcolor{qcv-trust}
& Agentic verify-retry loop
& \xmark & \xmark & \xmark & \xmark & \xmark & \xmark & \xmark & \xmark & \xmark & \pmark & \cmark \\
\cellcolor{qcv-trust}
& Production monitoring
& \xmark & \xmark & \xmark & \xmark & \xmark & \xmark & \xmark & \xmark & \xmark & \xmark & \cmark$^\P$ \\

\bottomrule[1.2pt]
\end{tabular}

\vspace{4pt}
\scriptsize
\textbf{Legend:}~\cmark~= fully supported;~\pmark~= partially supported;~\xmark~= not supported. \\
$^\dagger$Verilog HDL; $^\ddagger$General-domain code; $^\star$Unitary fidelity $\geq$0.99;
$^\diamond$Quantum simulator oracle; $^\S$Inference cost only; $^\P$Structured CSV logs. \\
QCalEval~\cite{cao2026qcaleval} = quantum calibration plot benchmark (243 samples, 87 scenario types, 22 experiment families). \\
\textit{Vis.$\to$Code} = MMCode~\cite{li2024mmcode}, Plot2Code~\cite{wu2025plot2code},
ChartCoder~\cite{zhao2025chartcoder}, CODE-VISION~\cite{wang2025codevision}.
\end{table*}

%% file: figures/workflow.tex
\begin{figure*}[!htbp]
\centering
\begin{tikzpicture}[node distance=0.5cm, every node/.style={font=\small},
  dbox/.style={draw, rounded corners, minimum height=0.7cm, minimum width=1.6cm, align=center, fill=orange!10},
  box/.style={draw, rounded corners, minimum height=0.8cm, minimum width=2.0cm, align=center, fill=blue!8},
  vbox/.style={draw, rounded corners, minimum height=0.8cm, minimum width=1.5cm, align=center, fill=green!8},
  arr/.style={-{Stealth[length=2.5mm]}, thick},
  darr/.style={-{Stealth[length=2mm]}, thick, dashed, gray}]
\node[draw, rounded corners, fill=orange!5, minimum width=14.0cm, minimum height=1.6cm] (dataset) at (6.5,3.2) {};
\node[font=\normalsize\bfseries] at (6.5,4.3) {QCV-Dataset (132 circuits, 5+ modalities)};
\node[dbox] (d1) at (1.5,3.2) {Images\\[-2pt]\scriptsize(132 PNG)};
\node[dbox] (d2) at (3.5,3.2) {Braket Code\\[-2pt]\scriptsize(132 .py)};
\node[dbox] (d3) at (5.5,3.2) {Qiskit Code\\[-2pt]\scriptsize(132 .py)};
\node[dbox] (d4) at (7.5,3.2) {Sim.\ Results\\[-2pt]\scriptsize(132 JSON)};
\node[dbox] (d5) at (9.5,3.2) {Annotations\\[-2pt]\scriptsize(bilingual)};
\node[dbox] (d6) at (11.5,3.2) {Targets\\[-2pt]\scriptsize(NL desc.)};
\node[box] (img) at (0.5,0.8) {Circuit\\[-2pt]Diagram};
\node[box] (llm) at (2.9,0.8) {MLLM\\[-2pt](BV / TV)};
\node[box] (code) at (5.6,0.8) {Generated\\[-2pt]Code};
\node[vbox] (syn) at (7.9,0.8) {Syntax\\[-2pt]Check};
\node[vbox] (exe) at (10.0,0.8) {Exec\\[-2pt]Check};
\node[vbox] (fid) at (12.2,0.8) {Fidelity\\[-2pt]$\geq 0.99$};
\draw[arr] (img) -- (llm);
\draw[arr] (llm) -- (code);
\draw[arr] (code) -- (syn);
\draw[arr] (syn) -- (exe);
\draw[arr] (exe) -- (fid);
\draw[decorate, decoration={brace, amplitude=4pt, raise=2pt}] (syn.north west) -- (fid.north east) node[midway, above=6pt, font=\small\itshape] {Verification Pipeline};
\draw[darr] (d1.south) -- ++(0,-0.6) -| (img.north);
\draw[darr] (d4.south) -- ++(0,-0.3) -| (fid.north);
\end{tikzpicture}
\caption{QCV system overview. \textbf{Top:} QCV-Dataset provides 132 quantum circuits with five core data modalities---circuit images, executable code (Braket and Qiskit), simulation results, bilingual annotations, and annotated failure cases---plus supplementary target descriptions and equivalence pairs. \textbf{Bottom:} The evaluation pipeline processes a circuit diagram image through an MLLM under BV or TV mode, then verifies the generated code through syntax checking, execution, and unitary matrix fidelity ($F \geq 0.99$).}
\label{fig:workflow}
\end{figure*}

%% file: tabs/dataset2.tex
\definecolor{d-basic}{RGB}{214, 232, 255}
\definecolor{d-inter}{RGB}{204, 242, 220}
\definecolor{d-advan}{RGB}{255, 236, 204}
\definecolor{d-block}{RGB}{255, 214, 224}
\definecolor{d-gate}{RGB}{232, 218, 255}
\definecolor{d-qubit}{RGB}{230, 240, 250}
\definecolor{d-class}{RGB}{225, 245, 235}
\definecolor{d-varia}{RGB}{250, 240, 225}
\definecolor{d-error}{RGB}{255, 225, 225}
\definecolor{d-qml}{RGB}{224, 242, 254}
\definecolor{d-bext}{RGB}{255, 235, 215}
\definecolor{d-visua}{RGB}{240, 240, 240}
\definecolor{d-btc}{RGB}{255, 228, 230}

\newcommand{\dmark}{\textcolor{green!60!black}{\ding{51}}}

\begin{table*}[t]
\caption{QCV-Dataset: 132 Circuits across 13 Categories}
\label{tab:dataset}
\footnotesize
\setlength{\tabcolsep}{4pt}
\renewcommand{\arraystretch}{1.05}
\centering
\begin{tabular}{@{}>{\raggedright\arraybackslash}p{2.2cm} >{\raggedright\arraybackslash}p{2.8cm} c >{\centering\arraybackslash}p{0.8cm} c >{\raggedright\arraybackslash}p{4.0cm} c@{}}
\toprule[1.2pt]
\textbf{Category} & \textbf{Focus} & \textbf{N} & \textbf{Qubits} & \textbf{Depth} & \textbf{Example Circuits} & \textbf{Eval} \\
\midrule[0.8pt]
\cellcolor{d-basic}\textbf{Basic} & Fundamental gates & 5 & 1--3 & 1--3 & Hadamard, Bell, Toffoli & \dmark \\
\cellcolor{d-inter}\textbf{Intermediate} & Multi-gate composition & 10 & 2--4 & 2--6 & QFT, Grover, Teleport & \dmark \\
\cellcolor{d-advan}\textbf{Advanced} & Complete algorithms & 6 & 3--5 & 4--14 & QAOA, VQE, BV & \dmark \\
\cellcolor{d-block}\textbf{Blockchain} & Quantum protocols & 11 & 2--8 & 1--9 & BB84, VRF, Consensus & \dmark \\
\midrule[0.4pt]
\cellcolor{d-gate}\textbf{A: Gate Cov.} & All standard gate types & 15 & 1--3 & 1--8 & CRy, CCZ, iSWAP & \dmark \\
\cellcolor{d-qubit}\textbf{B: Qubit Sc.} & 4--10 qubit circuits & 12 & 4--10 & 4--10 & GHZ-10, QFT-5, Ring & \dmark \\
\cellcolor{d-class}\textbf{C: Classical} & Textbook algorithms & 15 & 2--4 & 3--8 & Deutsch-Jozsa, Simon, Shor & \dmark \\
\cellcolor{d-varia}\textbf{D: Variational} & NISQ-era ans\"{a}tze & 10 & 2--4 & 3--27 & HW-efficient, UCCSD & \dmark \\
\cellcolor{d-error}\textbf{E: Error Corr.} & QEC codes & 8 & 3--9 & 2--5 & Shor-9, Steane-7, Surface & \dmark \\
\cellcolor{d-qml}\textbf{F: Quantum ML} & QNN, QCNN, kernels & 10 & 2--8 & 1--13 & QCNN-8q, QGAN, Kernel & \dmark \\
\cellcolor{d-bext}\textbf{G: BC Ext.} & Crypto protocols & 8 & 3--6 & 2--7 & E91, Voting, Auction & \dmark \\
\cellcolor{d-visua}\textbf{H: Visual Var.} & Layout robustness & 10 & 2--4 & 1--11 & Barrier, Decomposed, Reversed & \dmark \\
\cellcolor{d-btc}\textbf{I: BTC Sec.} & Threats \& PQC defense & 12 & 4--7 & 3--12 & Shor vs ECDSA, Kyber & \dmark \\
\midrule[0.4pt]
\multicolumn{2}{@{}l}{\textbf{Total}} & \textbf{132} & \textbf{1--10} & \textbf{1--27} & & \\
\bottomrule[1.2pt]
\end{tabular}
\end{table*}

%% file: tabs/result2.tex

\definecolor{r-opus}{RGB}{255, 214, 224}     
\definecolor{r-sonnet}{RGB}{204, 242, 220}   
\definecolor{r-haiku}{RGB}{255, 236, 204}    
\definecolor{r-delta}{RGB}{240, 240, 240}    
\definecolor{r-best}{RGB}{21, 128, 61}       

\begin{table}[t]
\centering
\caption{Pass rate on the 21-circuit core benchmark ($n=5$ repeated trials, mean$\pm$std\,\%). Fidelity $\geq 0.99$. BV = Basic Vision; TV = Thinking Vision (chain-of-thought).}
\label{tab:results}
\footnotesize
\setlength{\tabcolsep}{6pt}
\renewcommand{\arraystretch}{1.15}
\begin{tabular}{@{}l >{\columncolor{r-opus}}c >{\columncolor{r-sonnet}}c >{\columncolor{r-haiku}}c@{}}
\toprule[1.2pt]
& \textbf{Opus 4.6} & \textbf{Sonnet 4.6} & \textbf{Haiku 4.5} \\
& \scriptsize\textcolor{gray}{2.20$\times$ cr/call} & \scriptsize\textcolor{gray}{1.30$\times$ cr/call} & \scriptsize\textcolor{gray}{0.40$\times$ cr/call} \\
\midrule[0.8pt]
BV & 85.7$\pm$6 & \textbf{\textcolor{r-best}{91.4$\pm$5}} & 48.6$\pm$4 \\
TV & 81.9$\pm$8 & 88.6$\pm$4 & 53.3$\pm$9 \\
\midrule[0.4pt]
\cellcolor{r-delta}CoT $\Delta$ & \cellcolor{r-delta}$-$3.8pp & \cellcolor{r-delta}$-$2.9pp & \cellcolor{r-delta}+4.8pp \\
\cellcolor{r-delta}$p$-value & \cellcolor{r-delta}.26 & \cellcolor{r-delta}.19 & \cellcolor{r-delta}.69 \\
\bottomrule[1.2pt]
\end{tabular}
\end{table}

%% file: tabs/full-benchmark2.tex
\definecolor{bestgreen}{RGB}{129, 199, 132}    
\definecolor{strongblue}{RGB}{187, 222, 251}     
\definecolor{goodwhite}{RGB}{255, 255, 255}      
\definecolor{modentyellow}{RGB}{255, 245, 157}   
\definecolor{weakred}{RGB}{255, 205, 210}        
\definecolor{headerblue}{RGB}{25, 55, 109}       
\definecolor{totalbg}{RGB}{240, 244, 248}        

\newcommand{\bestcell}[1]{\cellcolor{bestgreen}\textbf{#1}}

\newcommand{\perfcell}[1]{\perfcellinner#1\relax}
\def\perfcellinner#1\relax{%
  \ifnum#1=100
    \cellcolor{strongblue}\textbf{100\%}%
  \else\ifnum#1>74
    \cellcolor{strongblue}#1\%%
  \else\ifnum#1>59
    \cellcolor{goodwhite}#1\%%
  \else\ifnum#1>49
    \cellcolor{modentyellow}#1\%%
  \else
    \cellcolor{weakred}#1\%%
  \fi\fi\fi\fi
}

\begin{table}[!htbp]
\centering
\caption{Pass rate on 132-circuit benchmark by category (fidelity $\geq 0.99$, $n=1$). \colorbox{bestgreen}{\textbf{B}}=best-in-row; \colorbox{strongblue}{\strut~}$\geq$75\%; \fcolorbox{gray!50}{white}{\strut~}60--74\%; \colorbox{modentyellow}{\strut~}50--59\%; \colorbox{weakred}{\strut~}$<$50\%. CoT $\Delta$ = TV$-$BV. }
\label{tab:full-benchmark}
\footnotesize
\setlength{\tabcolsep}{3.5pt}
\setlength{\aboverulesep}{2pt}
\setlength{\belowrulesep}{2pt}
\begin{tabular}{@{}p{1.6cm} c r c c c c c c c c@{}}
\toprule
& & & & \multicolumn{2}{c}{\cellcolor{headerblue}\textcolor{white}{\textbf{Opus 4.6}}} & \multicolumn{2}{c}{\cellcolor{headerblue}\textcolor{white}{\textbf{Sonnet 4.6}}} & \multicolumn{2}{c}{\cellcolor{headerblue}\textcolor{white}{\textbf{Haiku 4.5}}} & \\
\cmidrule(lr){5-6} \cmidrule(lr){7-8} \cmidrule(lr){9-10}
\textbf{Category} & \textbf{Dif.} & \textbf{\#} & \textbf{Q} & \cellcolor{headerblue}\textcolor{white}{BV} & \cellcolor{headerblue}\textcolor{white}{TV} & \cellcolor{headerblue}\textcolor{white}{BV} & \cellcolor{headerblue}\textcolor{white}{TV} & \cellcolor{headerblue}\textcolor{white}{BV} & \cellcolor{headerblue}\textcolor{white}{TV} & \textbf{Bst} \\
\midrule
Demo & B & 5 & 1--3 & \bestcell{100\%} & \bestcell{100\%} & \bestcell{100\%} & \bestcell{100\%} & \perfcell{80} & \perfcell{40} & \bestcell{100} \\
Gate Coverage & B & 15 & 1--3 & \bestcell{86\%} & \perfcell{80} & \perfcell{80} & \perfcell{80} & \perfcell{53} & \perfcell{53} & \bestcell{86} \\
Intermediate & M & 10 & 2--4 & \bestcell{90\%} & \bestcell{90\%} & \bestcell{90\%} & \bestcell{90\%} & \perfcell{30} & \perfcell{60} & \bestcell{90} \\
Visual Variants & M & 10 & 2--4 & \bestcell{100\%} & \perfcell{90} & \bestcell{100\%} & \perfcell{80} & \perfcell{50} & \perfcell{30} & \bestcell{100} \\
Qubit Scaling & M & 12 & 4--10 & \perfcell{91} & \perfcell{91} & \bestcell{100\%} & \perfcell{91} & \perfcell{66} & \perfcell{66} & \bestcell{100} \\
Advanced Algo. & H & 6 & 3--5 & \bestcell{83\%} & \bestcell{83\%} & \bestcell{83\%} & \perfcell{50} & \perfcell{33} & \perfcell{50} & \bestcell{83} \\
Classical Algo. & H & 15 & 2--4 & \bestcell{66\%} & \bestcell{66\%} & \perfcell{60} & \bestcell{66\%} & \perfcell{26} & \perfcell{33} & \bestcell{66} \\
Variational & H & 10 & 2--4 & \perfcell{70} & \perfcell{70} & \perfcell{80} & \perfcell{80} & \perfcell{50} & \perfcell{50} & \bestcell{80} \\
Error Correction & H & 8 & 3--9 & \bestcell{75\%} & \bestcell{75\%} & \perfcell{62} & \perfcell{62} & \perfcell{12} & \perfcell{25} & \bestcell{75} \\
Quantum ML & H & 10 & 2--8 & \perfcell{70} & \perfcell{80} & \perfcell{90} & \bestcell{100\%} & \perfcell{60} & \perfcell{80} & \bestcell{100} \\
Blockchain & H & 11 & 2--8 & \perfcell{63} & \perfcell{54} & \bestcell{81\%} & \bestcell{81\%} & \perfcell{45} & \perfcell{45} & \bestcell{81} \\
Blockchain Ext. & H & 8 & 3--6 & \bestcell{87\%} & \bestcell{87\%} & \perfcell{62} & \perfcell{62} & \perfcell{50} & \perfcell{62} & \bestcell{87} \\
BTC/Security & H & 12 & 4--7 & \bestcell{58\%} & \perfcell{41} & \perfcell{33} & \perfcell{33} & \perfcell{16} & \perfcell{16} & \bestcell{58} \\
\midrule
\rowcolor{totalbg}
\textbf{Total} & & \textbf{132} & \textbf{1--10} & \bestcell{78\%} & \perfcell{75} & \perfcell{77} & \perfcell{75} & \perfcell{43} & \perfcell{46} & \bestcell{78} \\
\textit{CoT} $\Delta$ & & & & \multicolumn{2}{c}{\textcolor{red!70!black}{$\blacktriangledown$\,$-$4}} & \multicolumn{2}{c}{\textcolor{red!70!black}{$\blacktriangledown$\,$-$3}} & \multicolumn{2}{c}{\textcolor{bestgreen!80!black}{$\blacktriangle$\,$+$5}} & \\
\bottomrule
\addlinespace[1pt]
\multicolumn{11}{@{}l@{}}{\tiny D=Difficulty (B=Basic, M=Med., H=Hard); Q=Qubits; Bst=Best(\%); BV=Basic Vision; TV=Thinking Vision (CoT).}
\end{tabular}
\end{table}

%% file: tabs/cost2.tex

\definecolor{c-haiku}{RGB}{255, 248, 225}   
\definecolor{c-sonnet}{RGB}{220, 252, 231}  
\definecolor{c-opus}{RGB}{255, 228, 230}    
\definecolor{c-header}{RGB}{245, 245, 245}  
\definecolor{c-mc-good}{RGB}{21, 128, 61}   
\definecolor{c-mc-bad}{RGB}{220, 38, 38}    

\begin{table}[t]
\centering
\caption{Cost-accuracy tradeoff (BV mode, 132 circuits). Credits are platform billing units; tiers reflect vendor capability-based pricing. MC = marginal cost per 1pp accuracy gain over the next-cheaper model.}
\label{tab:cost}
\footnotesize
\setlength{\tabcolsep}{5pt}
\renewcommand{\arraystretch}{1.15}
\begin{tabular}{@{}l c c c c c >{\raggedleft\arraybackslash}p{1.1cm}@{}}
\toprule[1.2pt]
\rowcolor{c-header}
\textbf{Model} & \textbf{Tier} & \textbf{Cr/call} & \textbf{Time} & \textbf{Pass\%} & \textbf{Cr/corr.} & \textbf{MC} \\
\midrule[0.8pt]
\cellcolor{c-haiku}Haiku 4.5 & \cellcolor{c-haiku}0.40$\times$ & \cellcolor{c-haiku}0.031 & \cellcolor{c-haiku}3.7s & \cellcolor{c-haiku}43\% & \cellcolor{c-haiku}0.072 & \cellcolor{c-haiku}--- \\
\cellcolor{c-sonnet}Sonnet 4.6 & \cellcolor{c-sonnet}1.30$\times$ & \cellcolor{c-sonnet}0.110 & \cellcolor{c-sonnet}6.0s & \cellcolor{c-sonnet}77\% & \cellcolor{c-sonnet}0.142 & \cellcolor{c-sonnet}\textbf{\textcolor{c-mc-good}{0.002}} \\
\cellcolor{c-opus}Opus 4.6 & \cellcolor{c-opus}2.20$\times$ & \cellcolor{c-opus}0.618 & \cellcolor{c-opus}24.4s & \cellcolor{c-opus}78\% & \cellcolor{c-opus}0.778 & \cellcolor{c-opus}\textcolor{c-mc-bad}{0.508} \\
\bottomrule[1.2pt]
\end{tabular}
\end{table}